\documentclass[a4paper,11pt]{article}
\pdfoutput=1 
\usepackage{lineno}
\usepackage{todonotes}
\usepackage{siunitx}
\usepackage{comment}
\usepackage[normalem]{ulem}
\usepackage{jinstpub} 
\usepackage{newtxtext,newtxmath}              

\title{\boldmath Intrinsic timing properties of ideal 3D-trench silicon sensor with fast front-end electronics}

\author[a,1]{Gian~Matteo~Cossu\note{Corresponding authors},}
\author[a,1]{Davide~Brundu}
\author[a]{and Adriano~Lai}

\affiliation[a]{INFN, Sezione di Cagliari, Cagliari, Italy}

\emailAdd{gianmatteo.cossu@ca.infn.it}

\abstract{
This paper describes the fundamental timing properties of a single-pixel sensor for charged particle detection
based on the 3D-trench silicon structure. We derive the results both analytically and numerically by considering a simple ideal sensor and the corresponding fast front-end electronics in two different case scenarios: ideal integrator and real fast electronics (trans-impedance amplifier). The particular shape of the Time of Arrival (TOA) distribution is examined and the relation between the time resolution and the spread of intrinsic charge collection time is discussed, by varying electronics parameters and discrimination thresholds. The results are obtained with and without simulated electronics noise. We show that the 3D-trench sensors are characterized by a \emph{synchronous region}, i.e. a portion of the active volume which leads to the same TOA values when charged particles cross it. The synchronous region size is dependent on the front-end electronics and discrimination threshold, and the phenomenon represents an intrinsic physical effect that leads to the excellent time resolution of these sensors. Moreover, we show that the TOA distribution is characterized by an intrinsic asymmetry, due to the 3D geometry only, that becomes negligible in case of significant electronics jitter.
}


\arxivnumber{2301.11190} 

\begin{document}
\maketitle
\flushbottom
\newpage

\section{Introduction}
High-luminosity conditions expected for the next generation of particle physics experiments require the use of vertex and tracking detectors with extreme requirements in terms of radiation resistance, spatial and time resolutions~\cite{CMS-TL,ATLAS-TL,LHCb-PII-Physics,FCC}. The renewed interest within the scientific community has led to several studies on different new technologies for future tracking systems, investigating also the possibility to measure simultaneously time and spatial coordinates (4D tracking)~\cite{PELLEGRINI201412,Parker,Hiroshima}. A comprehensive discussion on the topic can be found in ~\cite{CARTIGLIA2022167228,Lai_review}.

In this work, we study silicon 3D sensors~\cite{Parker} with trench structure, which have proven to be extremely radiation resistant and to have unprecedented time resolution, with the drawback of higher capacitance and complex fabrication~\cite{Brundu_2021,3Dradhard,Medoncino}. Beyond the experimental and technological aspects, an analytical study of their fundamental physical properties is not present in the literature. On the contrary, in-depth studies have been carried out on the physical properties of planar structures~\cite{riegler}. 
In this work we derive analytically the fundamental and intrinsic timing properties of 3D silicon sensors with trench technology in the case of ideal fast front-end electronics, starting from first principles.
In fact, it is well-known that, for planar sensors, signal shape variations as a function of the hit position in the pixel are caused by a non-uniform weighting field and non-saturated drift velocity, while here we focus on the fact that for silicon sensors with 3D-trench structure another source of signal shape variations is due to the different inducing times of charge carriers, which arise only from the particular 3D geometry.
Moreover, we discuss the results by using numerical simulations with custom simulation packages~\cite{tfboost_tcode}, in the case of realistic fast front-end electronics. All the results obtained both analytically and numerically are first discussed without introducing electronics noise (Sec.~\ref{sec:integrator}), while in Sec.~\ref{subsec:noise} we study the timing properties in the presence of noise. These results are in agreement with already performed experimental measurements on 3D-trench silicon pixel sensors, obtained at test beam~\cite{Brundu_2021,Testbeam10ps,Lampis:2022lpj} and with in-laboratory characterization with a laser setup~\cite{LaserIEEE}.


\section{Time resolution of ideal 3D-trench sensor with ideal fast electronics}
\label{sec:integrator}

In this section we discuss the transient currents in a 3D-trench silicon sensor, starting from first principles and deduce the properties of the voltage signal and the time of arrival (TOA) distribution by considering a transfer function of an ideal fast electronics (i.e. an ideal integrator).

\subsection{Properties of transient currents}

\begin{figure}[h]
	\centering
	\includegraphics[scale=0.25]{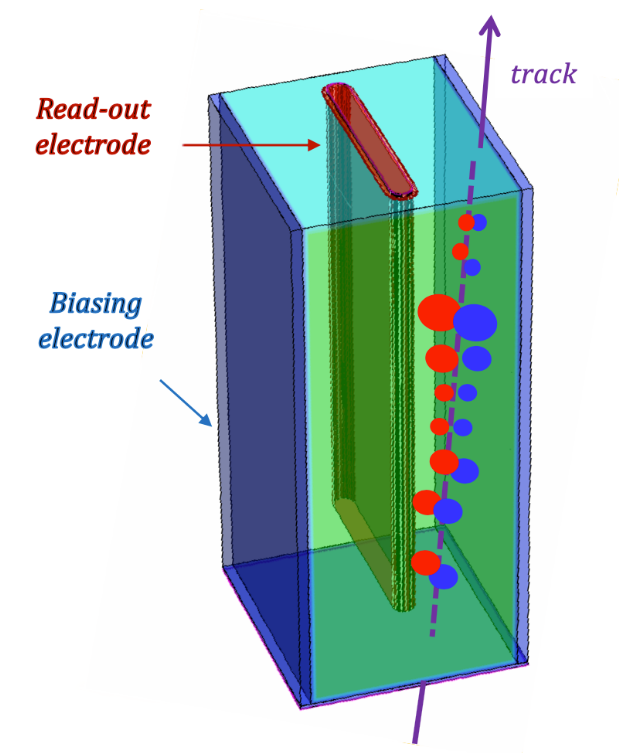}
	\caption{\footnotesize{Silicon sensor with 3D geometry, with track and charge deposit representation}.}
	\label{fig:general_3D}
\end{figure}

\begin{figure}[h]
	\centering
	\includegraphics[scale=0.4]{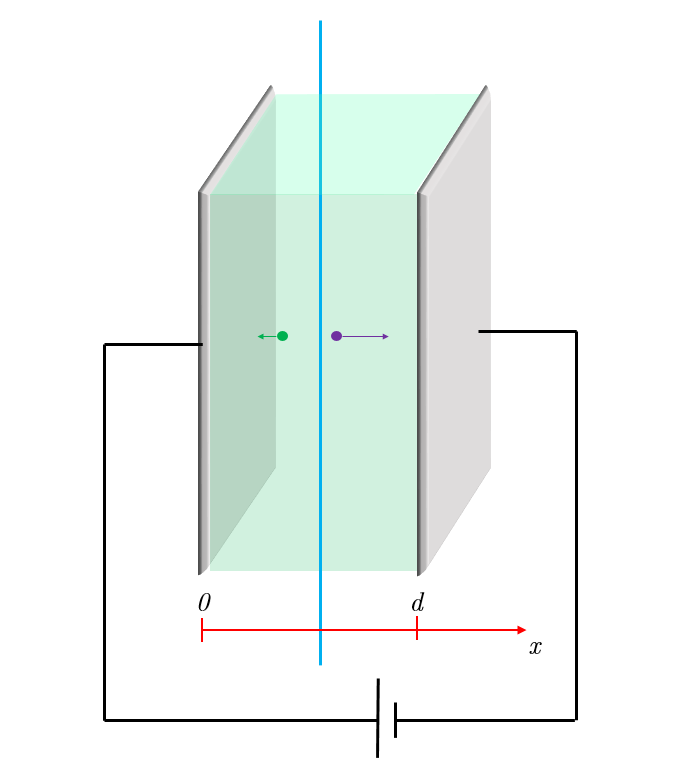}
	\includegraphics[scale=0.35]{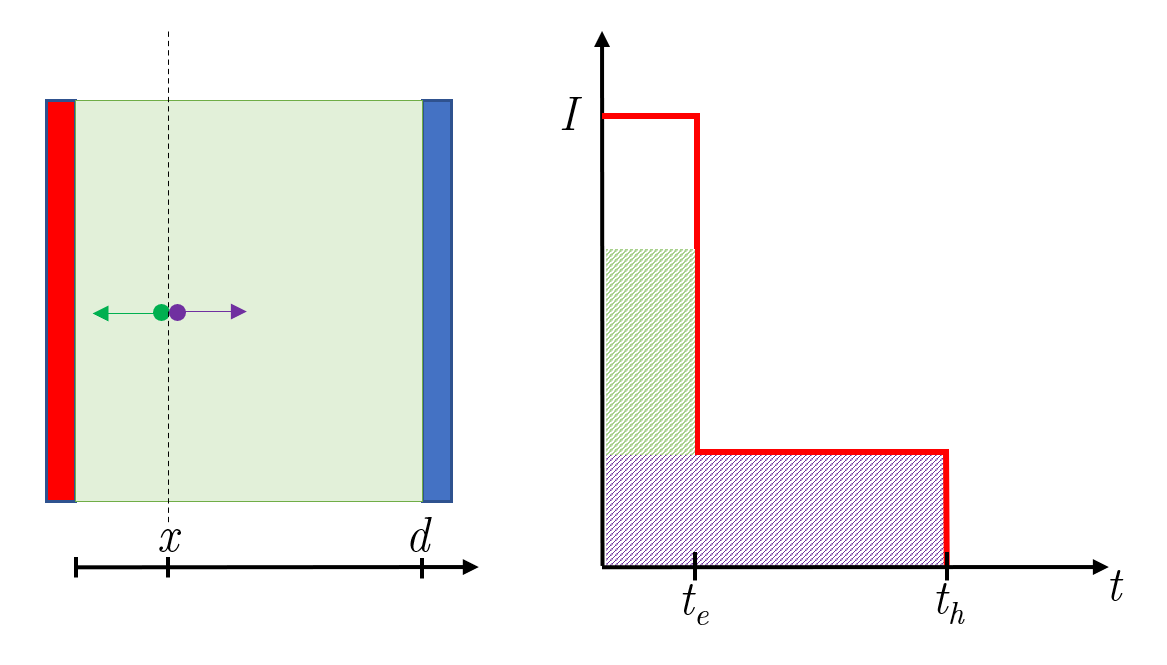}
	\caption{\footnotesize{Ideal parallel-plate sensor with 3D geometry (left), example of boxed current with induction duration $t_e$ for the electrons and $t_h$ for the holes (right)}}
	\label{fig:ideal3D}
\end{figure}

We consider a 3D silicon sensor with a standard parallel electrode configuration: two ohmic trenches are located at
the two opposite sides of the pixel, while a third trench, used as a readout electrode, is placed at the pixel centre, parallel to the two ohmic trenches (Fig.~\ref{fig:general_3D}). Without loss of generality, we can study the transient currents in one half of the sensor, by considering the region between the readout and one ohmic trench: the geometry is then a simple parallel-plate sensor, as shown in Fig.~\ref{fig:ideal3D}. 

In the case of pure vertical minimum ionizing particles (MIPs), the effect of Landau fluctuation of energy loss influences only signal amplitudes, causing a time walk effect on output signals. However, as we will discuss later, since this effect can be corrected by the proper discrimination method (constant fraction, use of time-over-threshold, etc.), we simplify the model by not introducing Landau fluctuations. In the case of particles crossing the sensor with a non-null tilt angle this approximation is, in principle, no more valid, since also the shape of the signals is affected. Moreover, we will not introduce higher-order effects like weighting field unevenness, non-saturated drift velocity, charge diffusion, etc. in order to derive time resolution contributions due to the 3D geometry only.

We indicate with $d$ the distance between electrodes and with $x$ the position of the vertical track passing through the sensor. 
By applying the Ramo theorem \cite{Ramo} to the configuration just discussed, in the case of tracks parallel to the electrodes, the resulting transient currents are simply boxed currents. Since all the carriers with the same polarity have always the same coordinate at every instant, all the electrons and holes currents can be summed up and the current is equivalent to the sum of two contributions,
\begin{equation}\label{curr}
I(t)= \frac{Q_{in}v_e}{d} \theta \Big(\frac{x}{v_e}-t \Big) + \frac{Q_{in}v_h}{d} \theta \Big(\frac{d-x}{v_h}-t \Big),
\end{equation}

where $Q_{\text{in}}$ is the overall charge deposited within the sensor, $v_e$ and $v_h$ are the carriers velocities, $t$ is the time and $\theta$ is the Heaviside step function. The duration of the signal is given by the carrier that has to cover the longer distance, depending on the $x$ position. We can define the two times, $t_e$ and $t_h$,
\begin{equation}\label{teth}
t_e=\frac{x}{v_e}, \hspace{1cm } t_h=\frac{d-x}{v_h}.
\end{equation}
The duration of the signal will be then the maximum between these two values,
\begin{equation}\label{tc_def}
t_c=\text{max} \{t_e,t_h\}.
\end{equation}
The center of gravity time (COG), as defined in \cite{riegler}, is
\begin{equation}\label{riegler}
{t_\text{cog}}(x)=\frac{1}{2d}\Big(\frac{x^2}{v_e}+\frac{(d-x)^2}{v_h}\Big).
\end{equation}

For electronics with a time constant $\tau\gg t_\text{cog}$ the time resolution can be estimated by considering the time displacement of the pulse response due to signals with COG time equal to $t_\text{cog}$. Following calculation in \cite{riegler_ppt} the standard deviation of the COG time can be written as

\begin{equation}\label{sigmatcog}
\sigma_{{t_\text{cog}}}=\sqrt{\overline{t_\text{cog}^2}-\overline{t_\text{cog}}^2}=\sqrt{\frac{4}{180}\frac{d^2}{v_e^2}-\frac{7}{180}\frac{d^2}{v_e v_h}+\frac{4}{180}\frac{d^2}{v_h^2}},
\end{equation}

where the over-line represents the average calculation. The resolution becomes smaller if the distance $d$ is kept smaller. It is worth noting that, with a planar detector without a gain mechanism, this leads to the fact that the noise will become the dominant contribution to time resolution~\cite{riegler}, not allowing to improve further the performance. This is not the case with 3D sensors, where the signal-to-noise ratio (SNR) can be kept high by choosing a proper thickness of the detector, allowing it to have more charge and, at the same time, maintaining the timing characteristics that depend only on the electrode distance.

\subsection{Properties of output signals and the synchronous region}

\begin{figure}[h]
	\centering
	\includegraphics[scale=0.35]{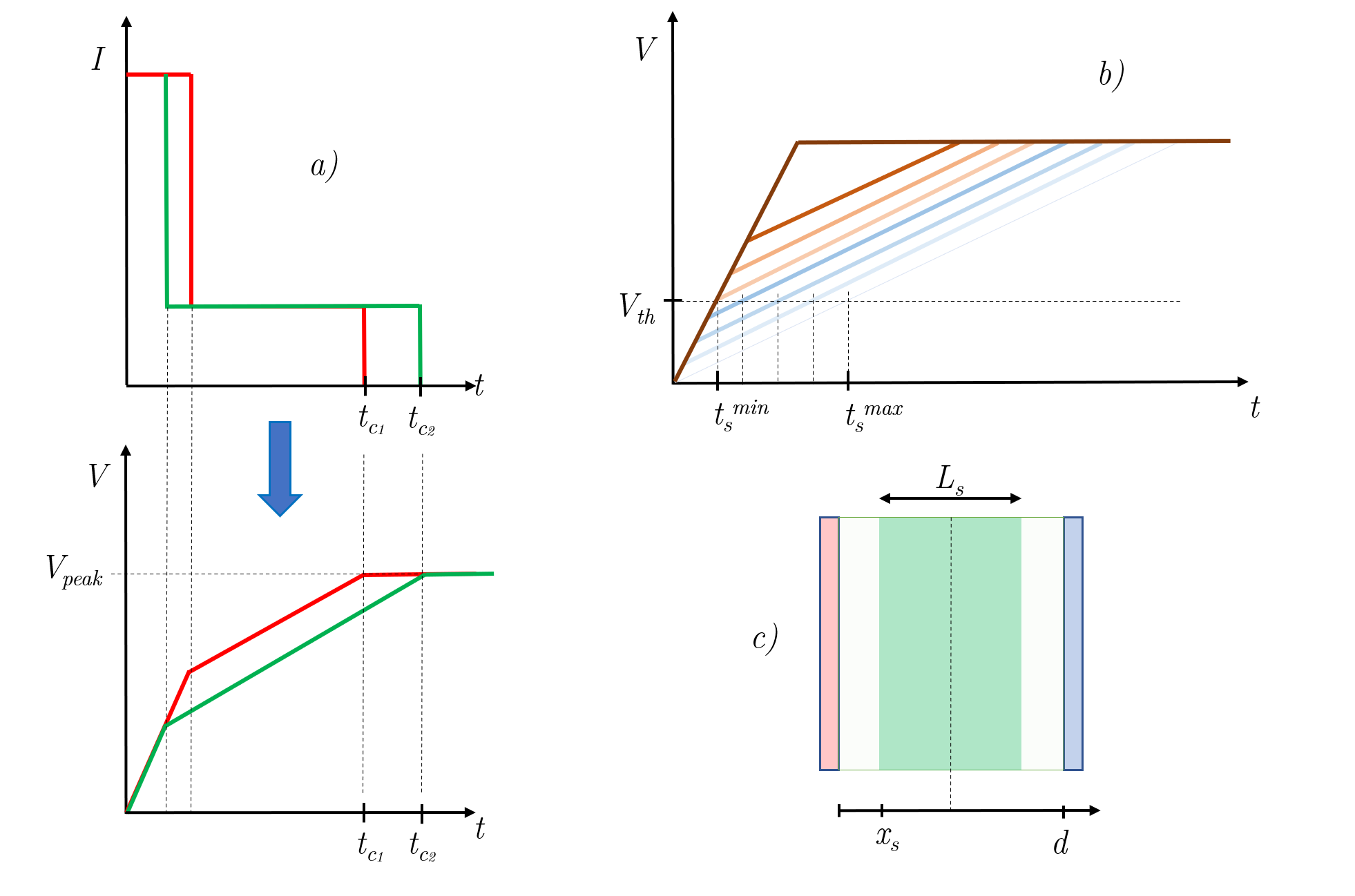}
	\caption{\footnotesize{a) Voltage signal for different boxed current, b) Different voltage signal and fixed threshold $V_{th}$: the brown signal will cross the threshold all at the same time $t_s^{min}$} while the blue signal will have different TOA ranging from $t_s^{min}$ to $t_s^{max}$, c) Synchronous region inside the detector (green region) with all events that produce the same time of arrival $t_s^{min}$}
	\label{fig:ideal3D2}
\end{figure}

In order to relate the intrinsic properties of currents with observed timing properties of 3D-trench sensors we need to perform a convolution between the transient currents in Eq.~\ref{curr} and a transfer function describing a fast front-end electronics. Let us consider a simple ideal integrator: the output voltage signal, $V(t)$, is the one obtained across the capacitance $C_D$ of the detector,
\begin{equation}
V(t)= \frac{1}{C_D}\int^{t}I(t') dt'.
\end{equation}
Considering the current in Eq.~\ref{curr}, the voltage signal for every position $x$ can be written as
\begin{equation}
V(t)=\frac{1}{C_D} \Bigg( \int_{0}^{t} \frac{Q_{in}v_e}{d} \theta \Big(\frac{x}{v_e}-t' \Big) dt' + \int_{0}^{t} \frac{Q_{in}v_h}{d} \theta \Big(\frac{d-x}{v_h}-t' \Big) dt'  \Bigg).
\end{equation}
Since the currents are boxed, with constant amplitude, $V(t)$ is a simple ramp signal. It is important to note that the slope  changes when one of the charge carriers stops inducing, namely when it has arrived at the corresponding electrode, as shown in Fig.~\ref{fig:ideal3D2}.

\vspace{0.2cm}
Given the $V(t)$ signal, let us focus now on the discrimination properties: if we consider a certain threshold, $V_{th}$, corresponding to a certain charge $Q_s=V_{th}C_D$, the threshold is reached at a time $t_s$ such that
\begin{equation}
V_{th}=V(t_s)=\frac{1}{C_D} \Bigg( \int_{0}^{t_s} \frac{Q_{in}v_e}{d} \theta \Big(\frac{x}{v_e}-t' \Big) dt' + \int_{0}^{t_s} \frac{Q_{in}v_h}{d} \theta \Big(\frac{d-x}{v_h}-t' \Big) dt'  \Bigg).
\end{equation}
Neglecting the capacitance $C_D$ for the moment, since it is just a multiplicative factor, the threshold charge $Q_s$ can be written as
\begin{equation}
Q_s=\Bigg( \int_{0}^{t_s} \frac{Q_{in}v_e}{d} \theta \Big(\frac{x}{v_e}-t' \Big) dt' + \int_{0}^{t_s} \frac{Q_{in}v_h}{d} \theta \Big(\frac{d-x}{v_h}-t' \Big) dt'  \Bigg),
\end{equation}
and by considering Eq.~\ref{teth} the previous equation becomes
\begin{equation}\label{Qst}
Q_s=\frac{Q_{in}v_e}{d} \Big[ \Big(t_s-t_e \Big)\theta \Big(t_e-t_s \Big) + t_e \Big] + \frac{Q_{in}v_h}{d} \Big[ \Big(t_s-t_h \Big)\theta \Big(t_h-t_s \Big) + t_h \Big].
\end{equation}

\begin{figure}[h]
	\centering
	\includegraphics[scale=0.30]{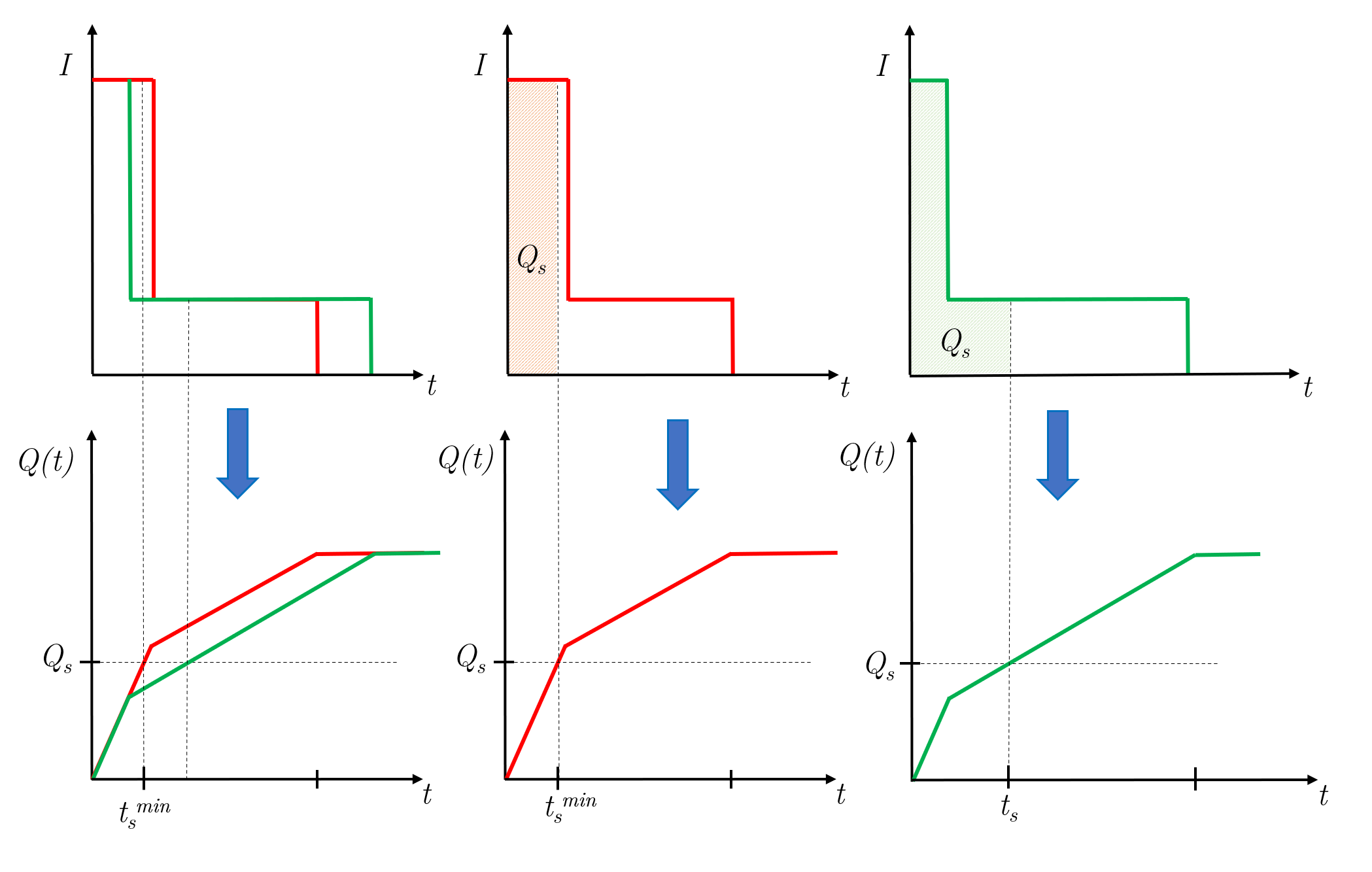}
	\caption{\footnotesize{Integrals of the boxed current: for the red current the charge $Q_s$ is reached with both carriers still contributing to the induction, for the green current one carrier has arrived at the corresponding electrode and the other has to induce the remaining charge to arrive at the value $Q_s$}.}
	\label{fig:ideal3D4}
\end{figure}

Depending on the impact point $x$ of the track, at the given threshold $V_{th}$, the time $t_s$ ranges between a minimum $t_s^{min}$ and a maximum $t_s^{max}$, as shown in Fig.~\ref{fig:ideal3D2} (b). It is interesting to examine the case of $t_s^{min}$, which is obtained by requiring the following conditions,
\begin{equation}\label{cond}
 t_s< t_e ~~~\text{and}~~~ t_s < t_h,
\end{equation}
which means that we are taking into account only the signals where the charge $Q_s$ is reached with \textbf{both carriers} still contributing to the induction, as shown in Fig.~\ref{fig:ideal3D2}~(b)  and Fig.~\ref{fig:ideal3D4}.

In this case, following Eq.~\ref{Qst}, we obtain that both the functions $\theta \big(t_e-t_s \big)$ and $\theta \big(t_h-t_s \big)$ are equal to 1 and we can write

\begin{equation}
Q_s=\frac{Q_{in}(v_e+v_h)}{d} t_s^{min},
\end{equation}
and, correspondingly,
\begin{equation}\label{min_ts}
t_s^{min}=\frac{Q_{s}d}{Q_{in}(v_e+v_h)}.
\end{equation}
This is an important result because the time $t_s^{min}$ depends only on constant values, having specified the threshold $Q_s$. Thus, it can be concluded that all the tracks impacting the sensor in positions $x$ satisfying the conditions in Eq.~\ref{cond} produce transient currents whose output signals reach the threshold at the same time $t_s^{min}$. This effect is peculiar of 3D-trench geometry, and the corresponding $x$ range defines a \textbf{Synchronous Region} within the sensor. The events belonging to the synchronous region satisfy
\begin{equation}\label{condition}
x= \begin{cases}  ~~~~x \geq v_e t_s  \hspace{1.5cm} ~~~~\text{for electrons,}\\ ~~~~x \leq d- v_h t_s \hspace{1.2cm} \text{for holes.} \end{cases}
\end{equation}

For simplicity, we can now consider the case where the velocities $v_e$ and $v_h$ are equal. The time becomes
\begin{equation} \label{mints}
t_s^{min}=\frac{d}{2v} \frac{Q_{s}}{Q_{in}} = \frac{d}{2v} \beta,
\end{equation}
where $\beta$ indicates the fraction of the threshold with respect to $Q_{in}$. The events satisfying Eq. \ref{condition} have positions $x \geq x_s$ and $x \leq d- x_s$ with
\begin{equation}
x_s=v t_s^{min},
\end{equation}
and stay in a region with width $L_s$ given by
\begin{equation}
L_s=d-2 x_s=d(1-\beta),
\end{equation}
which defines the width of the synchronous region, as shown in Fig.~\ref{fig:ideal3D2} (c). It is worth noting that $L_s$ depends on the chosen threshold $Q_s$.\\

On the contrary, events with impact position outside the synchronous region, $ x < x_s $ or $ x > d - x_s $, have a different time at threshold $t_s$, bigger than $t_s^{min}$. Given the symmetry of the problem, considering a position closer to the electrode where electrons induce and following Eq.~\ref{Qst} we obtain
\begin{equation}\label{qs_vevh}
Q_s=\frac{Q_{in}v_e}{d}t_e + \frac{Q_{in}v_h}{d}t_s,
\end{equation}
\begin{equation}
t_s=  \frac{d}{v} \frac{Q_{s}}{Q_{in}} - t_e,
\end{equation}
\begin{equation}
t_s=  2 t_s^{min} - \frac{x}{v}.
\end{equation}
We can see now that the time $t_s$ depends on the impact point $x$, i.e. the $x$ coordinates are defining a non-synchronous region. The maximum of $t_s$ is obtained for $x=0$, and peculiarly its value is twice $t_s^{min}$,
\begin{equation}
t_s^{max}= 2 t_s^{min}.
\end{equation}

\subsection{Time of arrival distribution and propagation coefficient}
\begin{figure}[h]
	\centering
	\includegraphics[scale=0.46]{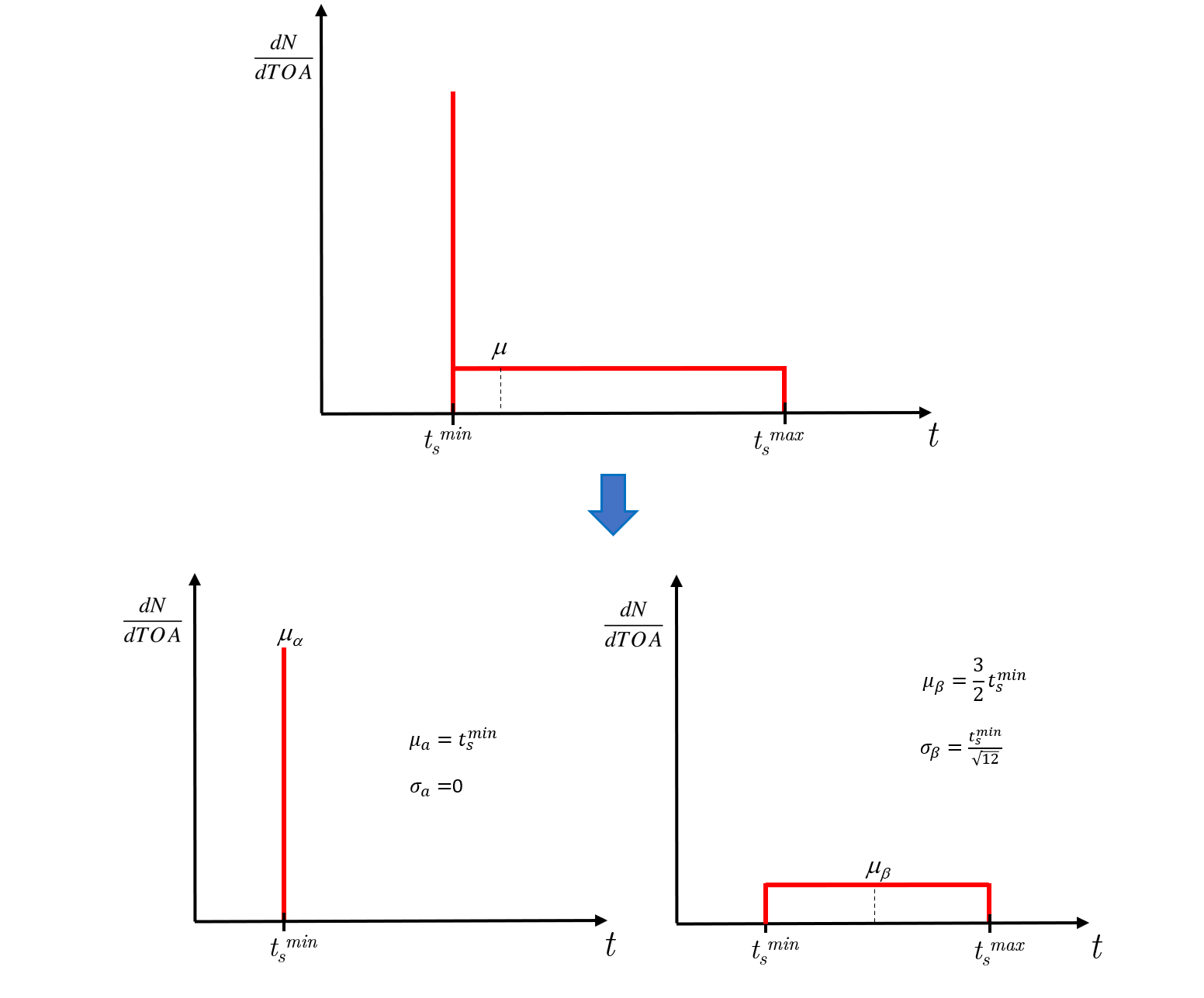}\vspace{-0.3cm}
	\caption{\footnotesize{ (Top) Distribution of time of arrival TOA with threshold $Q_s$: it presents a delta distribution with TOA $t_s^{min}$ for a fraction $\alpha$ of the events, and a uniform distribution due to fraction $\beta$ of events outside the synchronous region. (Bottom) Single distribution of the two regions: a delta distribution with $\sigma_{\alpha}=0$ and a uniform distribution with average $\mu_{\beta}$ and standard deviation $\sigma_{\beta}$ }.}
	\label{fig:distr}
\end{figure}

In order to study the time resolution of the ideal 3D-trench discussed above, we consider vertical tracks uniformly distributed on the coordinates $x$. All the tracks impacting the sensor within the synchronous region are characterized by a peaking TOA distribution because all the events have the same time of arrival $t_s^{min}$. A similar TOA distribution shape has been already seen in the case of Germanium gamma-ray detectors, where the sharp peak was due to events in
the central region of the detector \cite{GOULDING1966}. The events populating the narrow peak correspond to a fraction $\alpha$ of the total, given by
\begin{equation}
\alpha=\frac{Q_{in}-Q_s}{Q_{in}}.
\end{equation}
The remaining fraction $\beta=(1-\alpha$) shows a uniform TOA distribution (Fig.~\ref{fig:distr}) with width $\Delta t_s= t_s^{max}-t_s^{min}= t_s^{min}$ and standard deviation $\sigma_{\beta}= \frac{t_s^{min}}{\sqrt{12}}$.
The synchronous events, having all the same TOA, have a standard deviation
\begin{equation}
\sigma_{\alpha}= 0.
\end{equation}
We remark in this context that this result is ideal, with the aim of understanding the fundamental properties of 3D-trench geometry. Moreover, it is obtained by considering an ideal fast electronics without noise. As we will discuss later in the text, in the presence of noise the synchronous peak becomes larger due to the noise convolution, and the overall distribution is characterized by a narrow peak and a longer right tail, as observed experimentally and found in simulations \cite{Brundu_2021}.

The total TOA distribution is characterized by a probability density function (PDF) $p(t)$ as a mixture of two PDFs, indicated as $p_\alpha(t)$ and $p_\beta(t)$, for the synchronous and not synchronous contribution respectively. They are summed up with weights $\alpha$ and $\beta$ and are characterized by averages $\mu_{\alpha}=t_s^{min}$ and $\mu_{\beta}=\frac{3}{2}t_s^{min}$ and standard deviations $\sigma_{\beta}$ and $\sigma_{\alpha}$. The total PDF can be written as
\begin{equation}\label{mixturegeneral}
p(t) = \alpha  p_\alpha(t) + \beta p_\beta(t).
\end{equation}
The overall standard deviation of the mixture distribution is given by
\begin{equation}\label{mixture}
\sigma_{t_{s}}^2= \alpha(\sigma_{\alpha}^2+\mu_{\alpha}^2) + \beta(\sigma_{\beta}^2+\mu_{\beta}^2) - \mu^2,
\end{equation}
where $\mu= \alpha \mu_{\alpha}+ \beta \mu_{\beta}$,
\begin{equation}
\mu= (1-\beta) t_s^{min} + \beta \frac{3}{2}t_s^{min},
\end{equation}
\begin{equation}
\mu= t_s^{min} \frac{(2+\beta)}{2}.
\end{equation}
Substituting the corresponding values we find
\begin{equation}
\sigma_{t_{s}}^2= \frac{(t_s^{min})^2}{12}( 4\beta -3 \beta^2).
\end{equation}

\begin{figure}[h]
	\centering
	\includegraphics[scale=0.33]{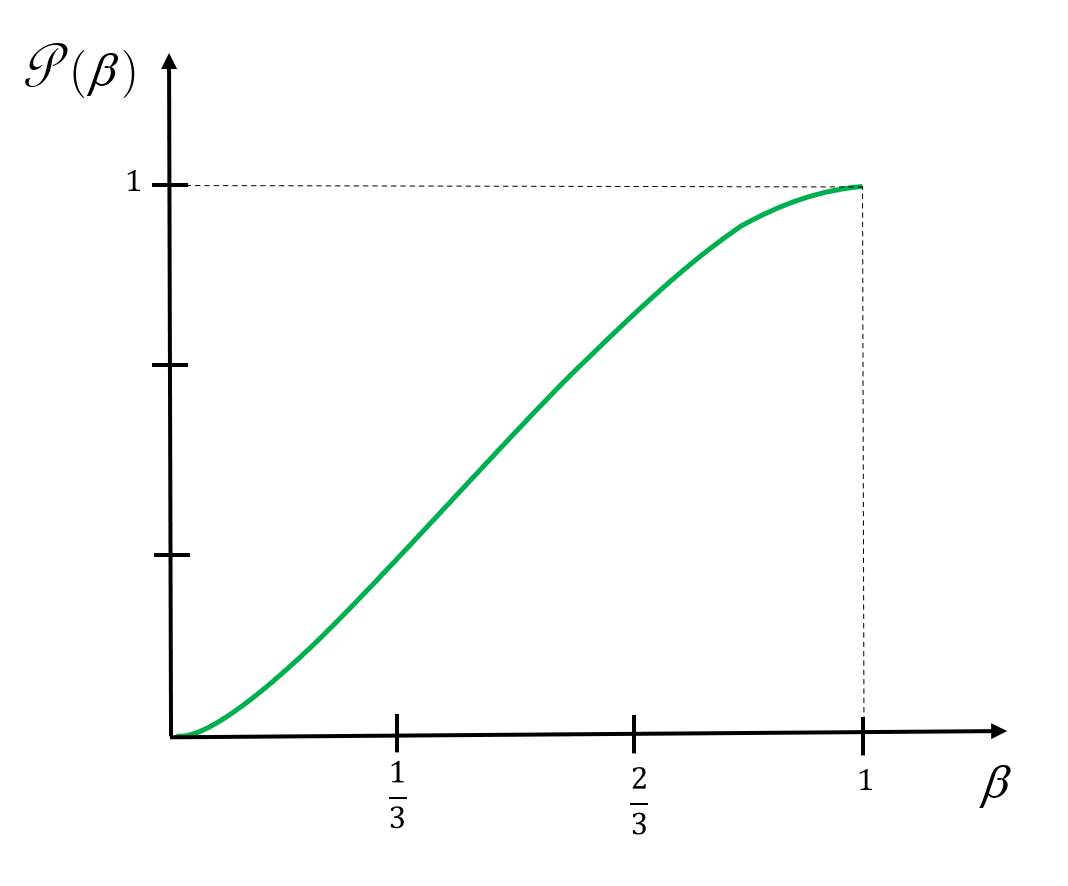}\vspace{-0.5cm}
	\caption{\footnotesize{ Propagation coefficient $\mathscr{P}(\beta)$ as a function of the threshold $\beta$.}}
	\label{falpha3}
\end{figure}

\begin{figure}[h]
	\centering
	\includegraphics[scale=0.46]{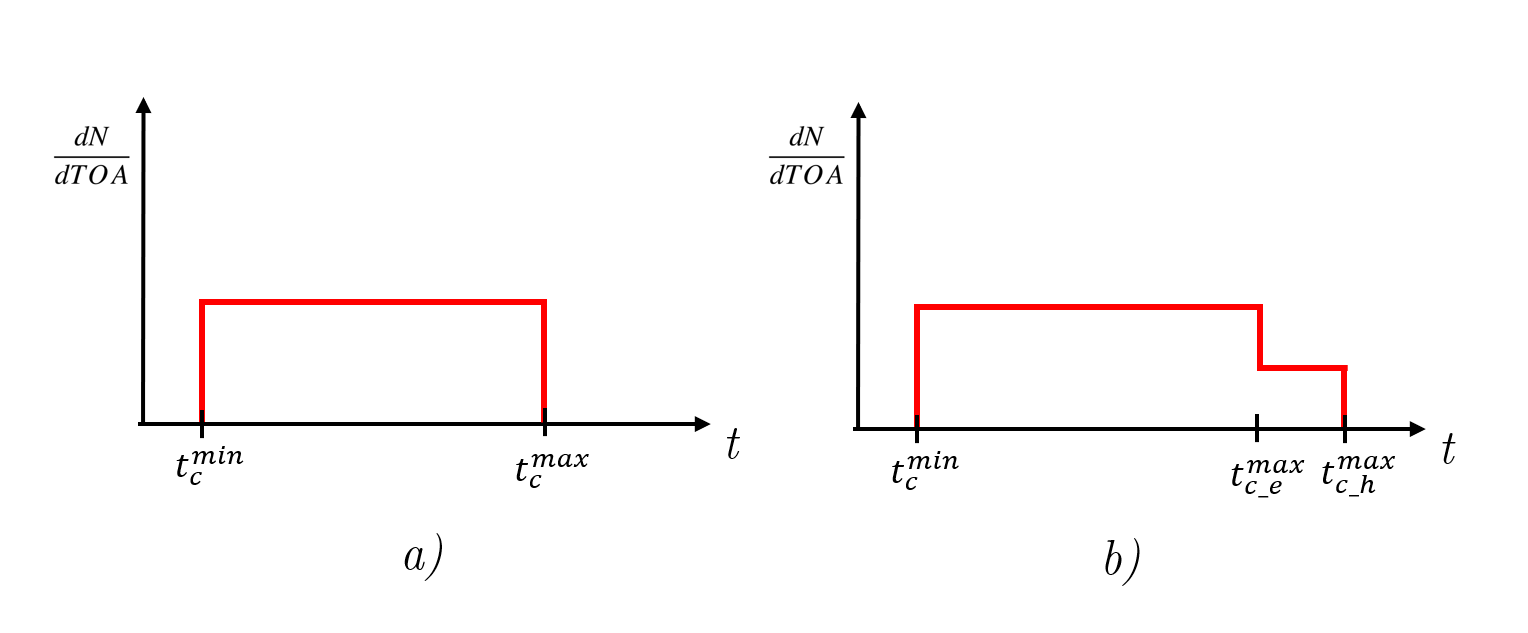}\vspace{-0.5cm}
	\caption{\footnotesize{ a) CCT distribution with same velocities $v$. b) CCT with different velocities $v_e$ and $v_h$.}}
	\label{CCT}
\end{figure}

Since the minimum time $t_s^{min}$ is proportional to the chosen threshold (Eq.~\ref{mints}), the total resolution can be also written as a function of the carrier velocity $v$, the electrode distance $d$ and the fraction $\beta$ as
\begin{equation}\label{risol}
\sigma_{t_s}=\frac{d}{2v} \frac{\beta \sqrt{4 \beta -3 \beta^2}}{\sqrt{12}},
\end{equation}
that is a monotonically increasing function with respect to the fraction $\beta$. At this point, it is interesting to relate the time resolution $\sigma_{t_s}$ of the output signal with the charge collection time (CCT) distribution, which is an intrinsic property of the sensor, namely the distribution populated with all the collection times $t_c$ as defined in Eq.~\ref{tc_def}. We find that for carriers with equal velocities this distribution is symmetric, as shown in Fig.~\ref{CCT} (a), with limits given by
\begin{equation}
t_c = \begin{cases}  t_c^{min}=\frac{d}{2v} & \mbox{if } x=\frac{d}{2}, \\ t_c^{max}=\frac{d}{v} & \mbox{if } x=d \end{cases},
\end{equation}
and, correspondingly, the standard deviation is given by
\begin{equation}\label{key}
\sigma_{t_c} = \frac{t_c^{max}-t_c^{min}}{\sqrt{12}} = \frac{t_c^{min}}{\sqrt{12}}.
\end{equation}

Thus, the time resolution in Eq.~\ref{risol} can be written as
\begin{equation}\label{prop_coeff}
\sigma_{t_s}={\beta \sqrt{4 \beta -3 \beta^2}} \cdot \sigma_{t_c} \implies \sigma_{t_s}=\mathscr{P}(\beta) \cdot \sigma_{t_c},
\end{equation}
where we introduced the function of the threshold $\mathscr{P}(\beta)$ called \textbf{Timing Propagation Coefficient} from the sensor to front-end electronics, as defined in \cite{FastTiming} and calculated analytically in this case for the ideal integrator (Fig.~\ref{falpha3}),
\begin{equation}\label{Pdibeta}
\mathscr{P}(\beta)={\beta \sqrt{4 \beta -3 \beta^2}}.
\end{equation}
The timing propagation coefficient measures the capability of fast electronics to reduce the intrinsic time collection dispersion of the sensor transient currents, as a function of the threshold chosen. In order to make a comparison between slow and fast electronics, let us consider the resolution calculated using Eq.~\ref{riegler} following \cite{riegler}, with equal carrier velocities $v$, 
\begin{equation}\label{tcog_equal}
\sigma_{t_{cog}}= \frac{d}{v}\frac{1}{\sqrt{180}},
\end{equation}
that can be written in terms of the standard deviation of the CCT $\sigma_{t_c}$ as
\begin{equation}\label{P_equalv}
\sigma_{t_{cog}}=  \frac{2}{\sqrt{15}} ~ \sigma_{t_c}.
\end{equation}
This calculation assumes electronics with time constant $\tau \gg t_{cog} $, thus for slow electronics and equal velocities the propagation coefficient $\mathscr{P}$ of an ideal 3D-trench silicon sensor is equal to
\begin{equation}
\mathscr{P}= \frac{2}{\sqrt{15}} \sim 0.516.
\end{equation}
and does not depend on the threshold. In our calculation, we find that a fast electronics, e.g an ideal integrator, reaches the same result if a threshold $\beta \sim 0.46$ is chosen, while for lower thresholds the resolution improves significantly (e.g. $\mathscr{P}=0.33$ for $\beta=0.33$ ).

\subsection{Asymmetry of the TOA distribution} \label{asymmetry}
\begin{figure}[h]
	\centering
	\includegraphics[width=\textwidth]{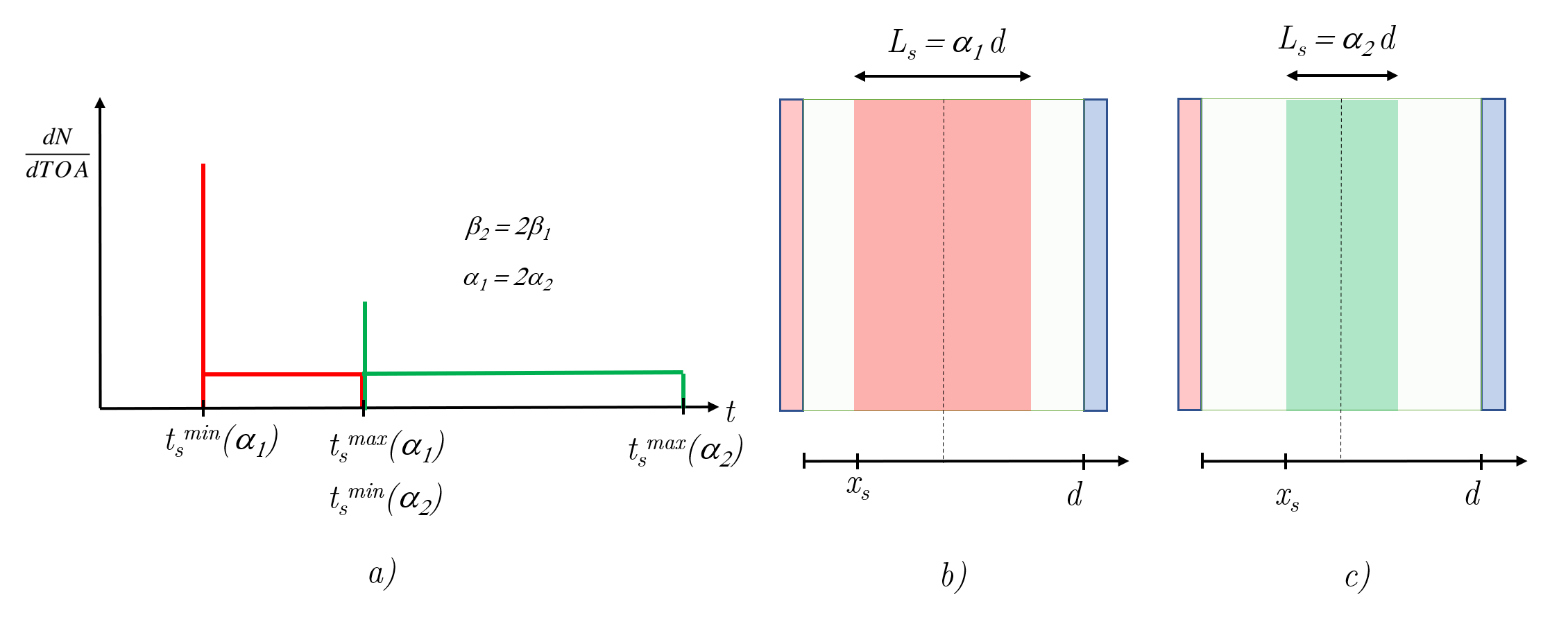}
	\caption{\footnotesize{ a) TOA distribution for different thresholds $\beta$. doubling the threshold, i.e $\beta_2=2 \beta_1$ the distribution gets larger and delayed; also the synchronous peak halves in amplitude. b) synchronous region for $\alpha_1$. c) synchronous region for $\alpha_2=\alpha_1 / 2$} }
	\label{fig:regsinc}
\end{figure}

Another important aspect of the TOA distribution is that the synchronous peak causes the distribution to become asymmetric, even in the case of carriers with the same saturation velocities $v$. Since we are studying an ideal sensor, with a constant electric field and constant saturation velocities of carriers over the entire active region of the sensor, this asymmetry cannot be caused by unevenness or non-uniformity of the weighting field. Indeed, it is an intrinsic characteristic of the 3D geometry and it is strictly related to the different charge collection times, since the charged carriers have to cover different distances depending on the position $x$. The \textbf{intrinsic asymmetry} becomes stronger if a lower threshold $\beta$ is chosen (higher $\alpha$) and the resolution improves (Fig. \ref{fig:regsinc}).

\begin{figure}[h]
	\centering
	\includegraphics[scale=0.32]{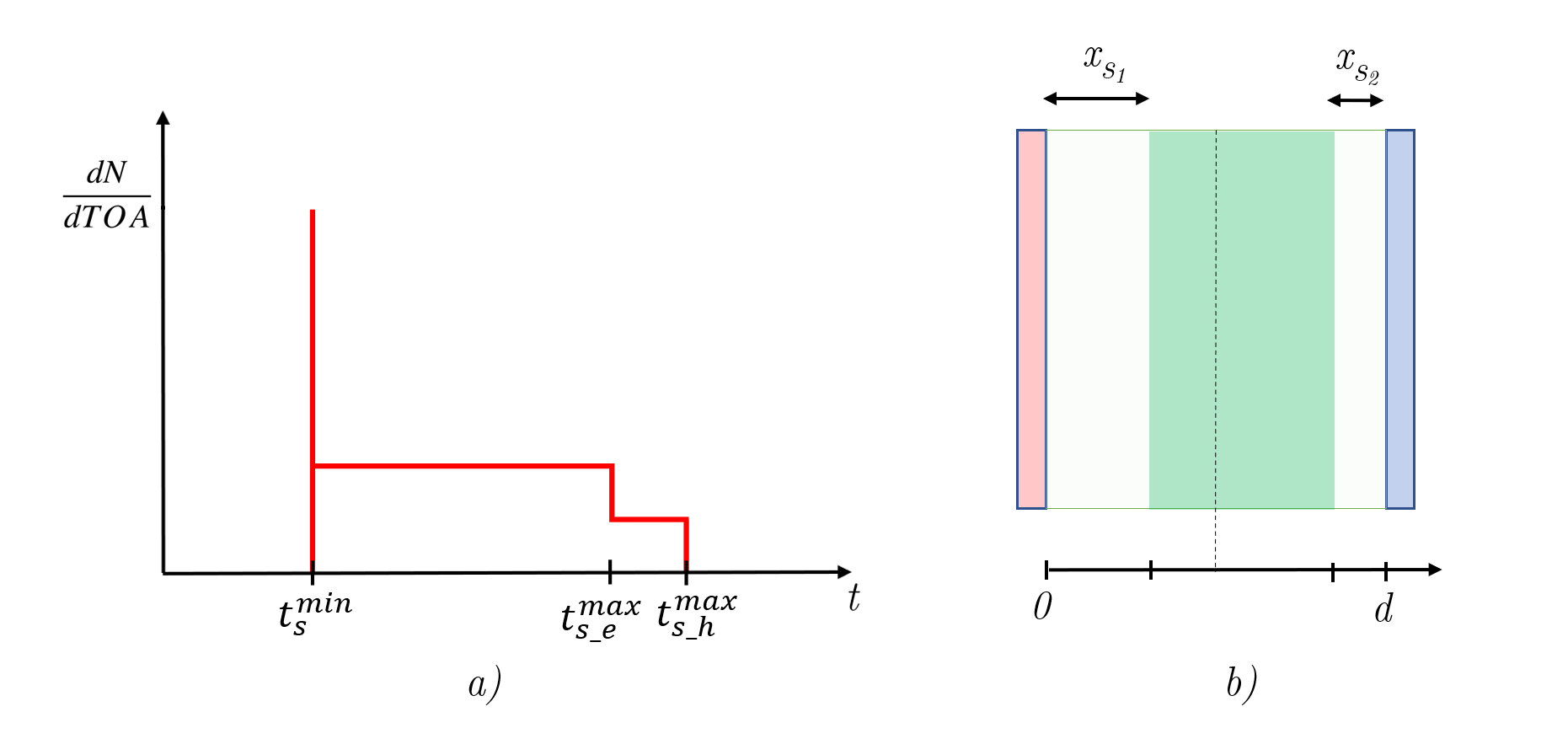}
	\caption{\footnotesize{ a) Time of arrival distribution in the case we have different saturation velocities $v_e$ and $v_h$. b) Synchronous region with different drift velocities for the carriers, that gets closer to the holes electrode.} }
	\label{fig:distr_2}
\end{figure}

 Indeed the size of the synchronous region becomes smaller with a higher threshold $\beta$ (Fig.~\ref{fig:regsinc}). We can generalize easily the results to the case of different saturation velocities. The TOA distribution acquires another source of asymmetry as shown in Fig.~\ref{fig:distr_2} (a). Following Eq.~\ref{qs_vevh}, we find two values for the time $t_s^{max}$, in particular
\begin{equation}
t_{s,e}^{max}= \frac{Q_s d}{Q_{in}v_e},~~~~~~~~t_{s,h}^{max}= \frac{Q_s d}{Q_{in}v_h}.
\end{equation}
The minimum value, $t_s^{min}$, is given by Eq.~\ref{min_ts}. In this case, the synchronous region is no more symmetric with respect to the sensor geometry, and it gets closer to the electrode at a lower potential, as shown in Fig.~\ref{fig:distr_2} (b). The size of the synchronous region is given by
\begin{equation}
L_s= d-x_{s_1}-x_{s_2},~~~~~~~~~~~~x_{s_1}=v_e t_s^{min},~~~~~~~~~~~~x_{s_2}=v_h t_s^{min}.
\end{equation}

\subsubsection{Propagation coefficient $\mathscr{P}$ with different saturation velocities}

The propagation coefficient $\mathscr{P}$ in case of different saturation velocities can be obtained by considering the CCT distribution shown in Fig.~\ref{CCT} (b).
The standard deviation of the total distribution can be calculated again by considering the mixture of two distributions, finding

\begin{equation}\label{sigmatc_vevh}
\sigma_{t_c}^2=\frac{(t_c^{min})^2}{12}\frac{v_e^4+2v_e^3 v_h -5v_e^2v_h^2+2v_ev_h^3+v_h^4}{v_e^2 v_h^2}.
\end{equation}

If we introduce the parameter $k$ as the ratio of hole and electron velocities, $v_h=k \cdot v_e$, we can express the standard deviation $\sigma_{t_{cog}}$ in Eq.~\ref{sigmatcog} in terms of $\sigma_{t_c}$,
\begin{equation}
\sigma_{t_{cog}}^2=\frac{(4k^2-7k+4)(1+k)^2 }{15(k^4+2k^3-5k^2+2k+1)} {\sigma_{t_c}^2}.
\end{equation}

\begin{figure}[h]
	\centering
	\includegraphics[width=\textwidth]{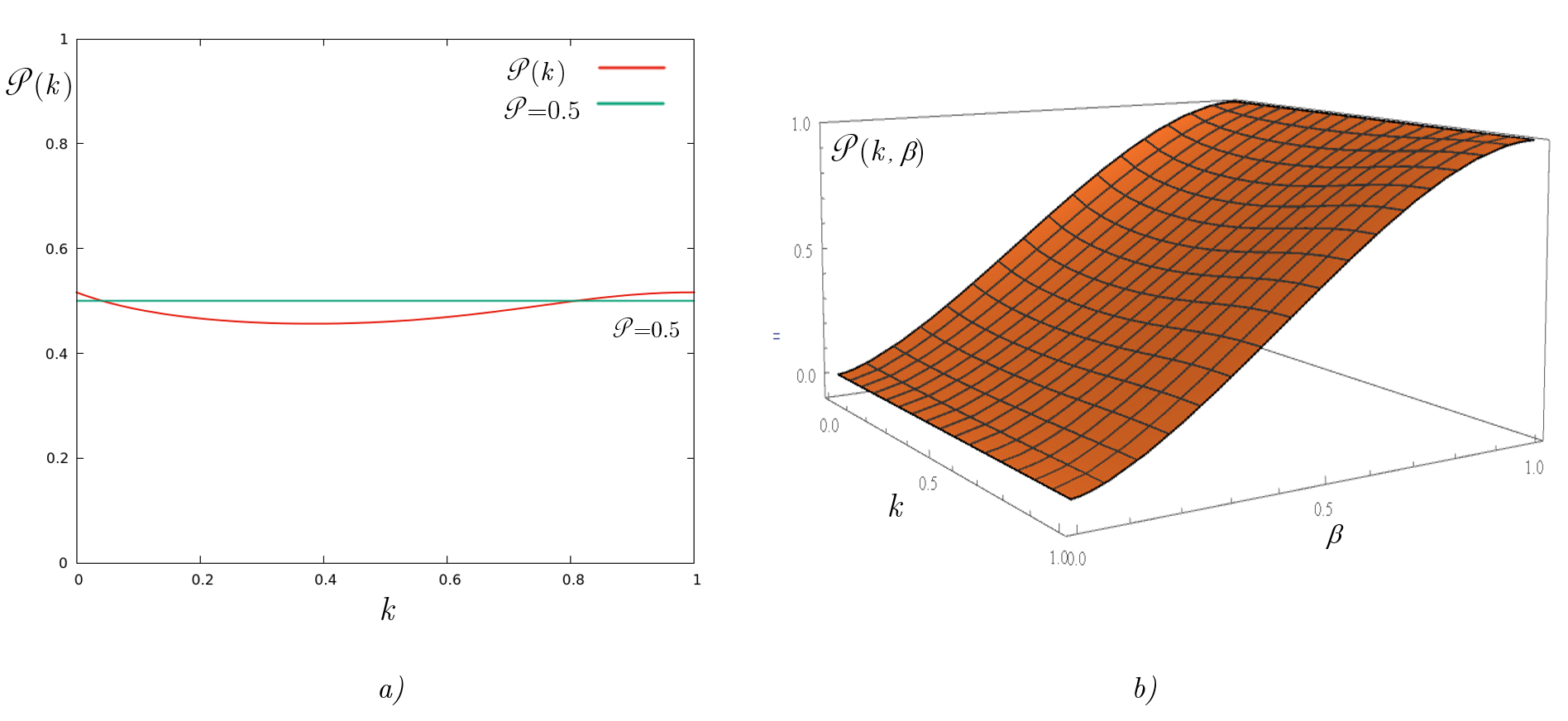}
	\caption{\footnotesize{ a) Timing propagation coefficient with slow electronics as a function of the ratio of the saturation velocities $k$. For equal saturation velocities, we get the value \ref{P_equalv}, for smaller $k$ (i.e smaller holes velocities) the propagation coefficient gets smaller reaching a minimum where $k=\frac{3-\sqrt{5}}{2}\sim 0.38$ (i.e $v_e\sim 2.63 ~v_h$). b) Timing propagation coefficient for the ideal integrator as a function of the threshold $\beta$ and the ratio $k$ of the velocities of the carriers. } }
	\label{fig:pdik}
\end{figure}

Thus, the following function,

\begin{equation}
\mathscr{P}(k)=\sqrt{\frac{(4k^2-7k+4)(1+k)^2 }{15(k^4+2k^3-5k^2+2k+1)}},
\end{equation}

represents the \textit{Timing propagation coefficient} of an ideal 3D-trench sensor with slow front-end electronics as a function of the factor $k$. The function is shown in Fig.~\ref{fig:pdik} (a). For equal velocities (i.e $k=1$) we obtain the propagation coefficient already discussed in Eq.~\ref{P_equalv}, while for $k<1$ the value decreases until reaches its minimum, given by
\begin{equation}
\mathscr{P}\Big(k=\frac{3-\sqrt{5}}{2}\Big)=\frac{\sqrt{30}}{12}\sim 0.456.
\end{equation}
This justifies a criterion that, in order to estimate the achievable resolution of a 3D silicon sensor that produces currents with different collection times $t_c$, one can consider the CCT distribution and take half the standard deviation $\sigma_{t_c}$. 

It is instructive to consider an even simpler model of sensor, where the currents are rectangular pulses with different duration $t_c$, uniform collection time distribution with standard deviation $\sigma_{t_c}$, following \cite{FastTiming}. In this case, for slow front-end electronics, we find that $\sigma_{t_s}$ is exactly half of $\sigma_{t_c}$, as the centroid resolution would suggest.

The definition of the \textit{timing propagation coefficient} becomes useful when we consider faster electronics because, as shown in \cite{FastTiming} and \ref{Pdibeta}, this can be expressed as a function of the threshold, to manifest the ability of a fast electronics to reduce the intrinsic collection time dispersion. Considering now the distribution Fig.~\ref{fig:distr_2}, we can take the mixture of three distributions, following the same method used in the previous sections, to find the propagation coefficient for the ideal integrator with different saturation velocities of carriers. In particular, we have
\begin{equation}
\sigma_{t_s}^2=\frac{\beta^3\Big(4(v_e^4-v_e^3v_h+v_e^2 v_h^2-v_ev_h^3+v_h^4)-3\beta(v_e^2-v_e v_h+v_h^2)^2\Big)}{(v_e^4+2v_e^3v_h-5v_e^2 v_h^2+2v_ev_h^3+v_h^4)}\sigma_{t_c}^2,
\end{equation}

\begin{equation}
\mathscr{P}(\beta,v_e,v_h)=\sqrt{\frac{\beta^3\Big(4(v_e^4-v_e^3v_h+v_e^2 v_h^2-v_ev_h^3+v_h^4)-3\beta(v_e^2-v_e v_h+v_h^2)^2\Big)}{(v_e^4+2v_e^3v_h-5v_e^2 v_h^2+2v_ev_h^3+v_h^4)}}.
\end{equation}

For $k=1$ we find again the propagation coefficient given by~\ref{Pdibeta}. The propagation coefficient for different drift velocities is shown in Fig.\ref{fig:pdik} (b).

\subsection{Correlation between time resolution and efficiency}
The existence of the synchronous region has the effect to produce a correlation between the resolution of the 3D detector and the efficiency.  In fact, once a certain threshold is chosen, if we are willing to sacrifice efficiency and consider only the events crossing the synchronous region, we would have a perfect detector with an ideal infinite resolution. This means that with the ideal integrator and a threshold $Q_s$ at $\beta=30\%$, as an example, the synchronous region would cover 70\% of the sensor active volume (i.e. not taking into account the trenches volume), thus obtaining a detector with an efficiency $E_{\text{ff}}=70\%$ and resolution $\sigma_{t_s}=0\,ps$. We remind that this result is derived by considering perpendicular tracks.
The calculation in the previous section can be extended to consider also the efficiency, thus we can define a new \textit{Propagation Coefficient}, $\mathscr{P}(\beta, E_{\text{ff}})$, as

\begin{equation}
\mathscr{P}(\beta, E_{\text{ff}})=\frac{\sqrt{(\beta + E_{\text{ff}}- 1)^3(3 + E_{\text{ff}}-3 \beta)}}{E_{\text{ff}}},
\end{equation}

\begin{equation}
\sigma_{t_s}=\mathscr{P}(\beta, E_{\text{ff}}) \cdot \sigma_{t_c}.
\end{equation}

\begin{figure}[h]
	\centering
	\includegraphics[scale=0.40]{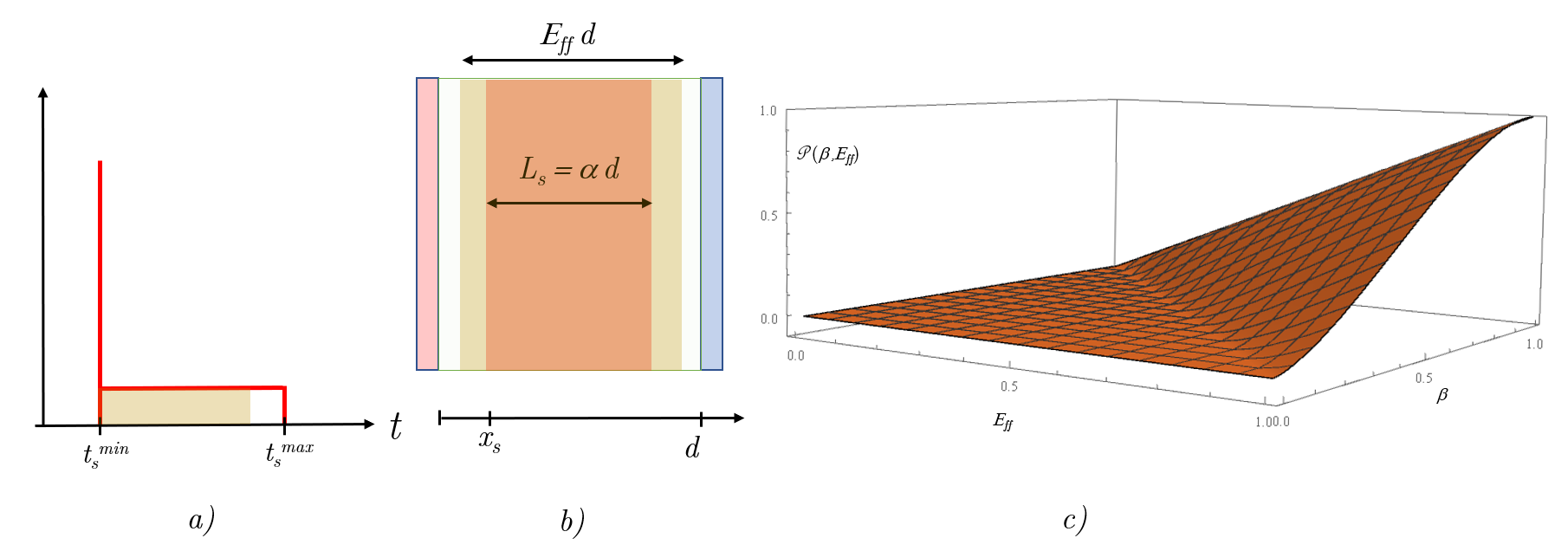}\vspace{-0.2cm}
	\caption{\footnotesize{ a) TOA Distribution for a chosen $\beta$ with highlighted the events for a reduced efficiency $E_{\text{ff}}$. b) Detector with a fraction $E_{\text{ff}}$ of the detector and the corresponding synchronous region $L_s$ inside. c) Propagation coefficient as a function of $\beta$ and $E_{\text{ff}}$: the flat region is where the propagation coefficient $\mathscr{P}(\beta, E_{\text{ff}}) =0$ and the detector is a perfect detector.} }
	\label{fig:Effi}
\end{figure}

\begin{figure}[h]
	\centering
	\includegraphics[scale=0.43]{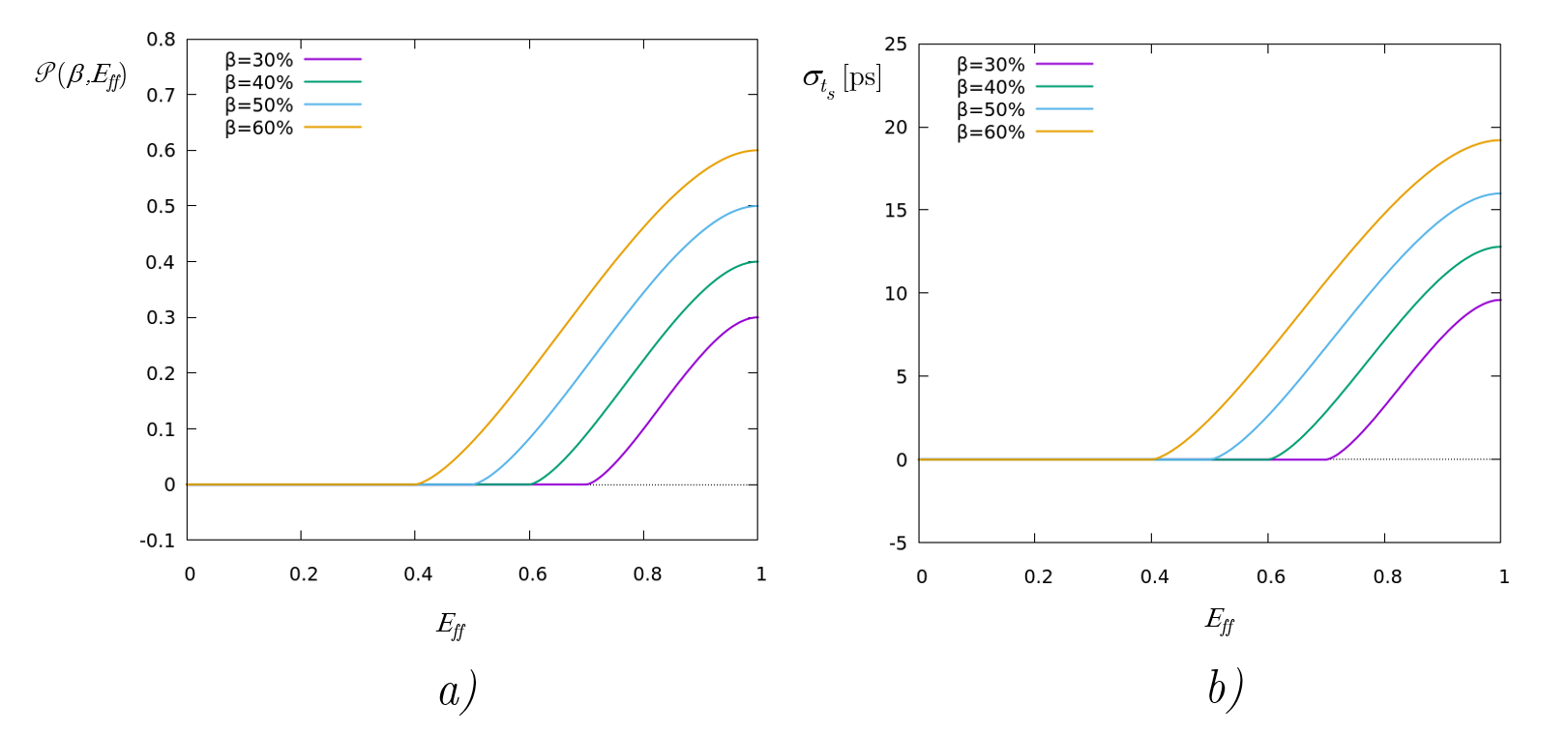}\vspace{-0.2cm}
	\caption{\footnotesize{ a) Propagation coefficient $\mathscr{P}(\beta, E_{\text{ff}})$ for different thresholds $\beta$. b) Resolution as a function of the efficiency $E_{\text{ff}}$ for different thresholds $\beta$ of an ideal 3D detector with $d=20~\mu m$ and saturation velocities $v=0.1 \mu m / ps$.} }
	\label{fig:Effi2}
\end{figure}

In this case, if an efficiency $E_{\text{ff}}$ equal or smaller than $\alpha$ is chosen (also $E_{\text{ff}}\leq 1-\beta$), the resolution becomes zero since we consider only events that are in the synchronous region (Fig. \ref{fig:Effi} and Fig. \ref{fig:Effi2} ).  If we increase the efficiency the resolution has to gradually worsen until the value in Eq. \ref{prop_coeff} is reached.

\newpage

\section{Time resolution of ideal 3D-trench sensor with real fast electronics}
\label{sec:tia}
In the previous section we discussed the properties of the voltage output signals and the corresponding TOA distribution by considering a transfer function of an ideal integrator. For a meaningful comparison with experimental observations, a more realistic transfer function should be considered. Unfortunately, the analytical calculations using such transfer functions are no more straightforward, so we will derive the results by using numerical simulations, with the possibility of interpreting the results in light of the analytical description carried out for the simpler model. 
The simulation of the front-end electronics has been performed by using the custom and open-source software called TFBoost~\cite{tfboost_tcode,TFBoost_cit}, capable of carrying out precise and very fast simulations of the entire read-out chain of 3D sensors~\cite{Brundu_2021}.

\subsection{Fast trans-impedance amplifier transfer function}

The model of the ideal integrator is useful to understand what happens when the electronics has a rise-time of the same order of magnitude as the current pulse. A more realistic electronics response can be modelled as a second-order trans-impedance amplifier (TIA), with DC trans-impedance $R_{m_0}$.  The linear system composed of sensor and electronics is characterized by a finite bandwidth and a time constant $\tau$. The transfer function can be written as
\begin{equation}
R_m(s) = \frac{R_{m_0}}{(1+s\tau)^2}.
\label{eq:RmFast}
\end{equation} 
The output signals with this type of fast electronics, when the time constant $\tau$ is of the same order of magnitude as the average current pulse duration, can be obtained with TFBoost by making the convolution between the transfer function in Eq.~\ref{eq:RmFast} and the boxed currents of the ideal 3D sensor previously discussed (Fig.~\ref{fig:ideal3D}). Within TFBoost the transfer function is described by using a small signal model, whose parameters are chosen in order to have $\tau=160$ ps and $R_{m_0}=2.3~\text{k}\Omega$, representing a realistic front-end electronics as described in \cite{FastTiming}. 

\begin{figure}[h]
	\centering
	\includegraphics[scale=0.39]{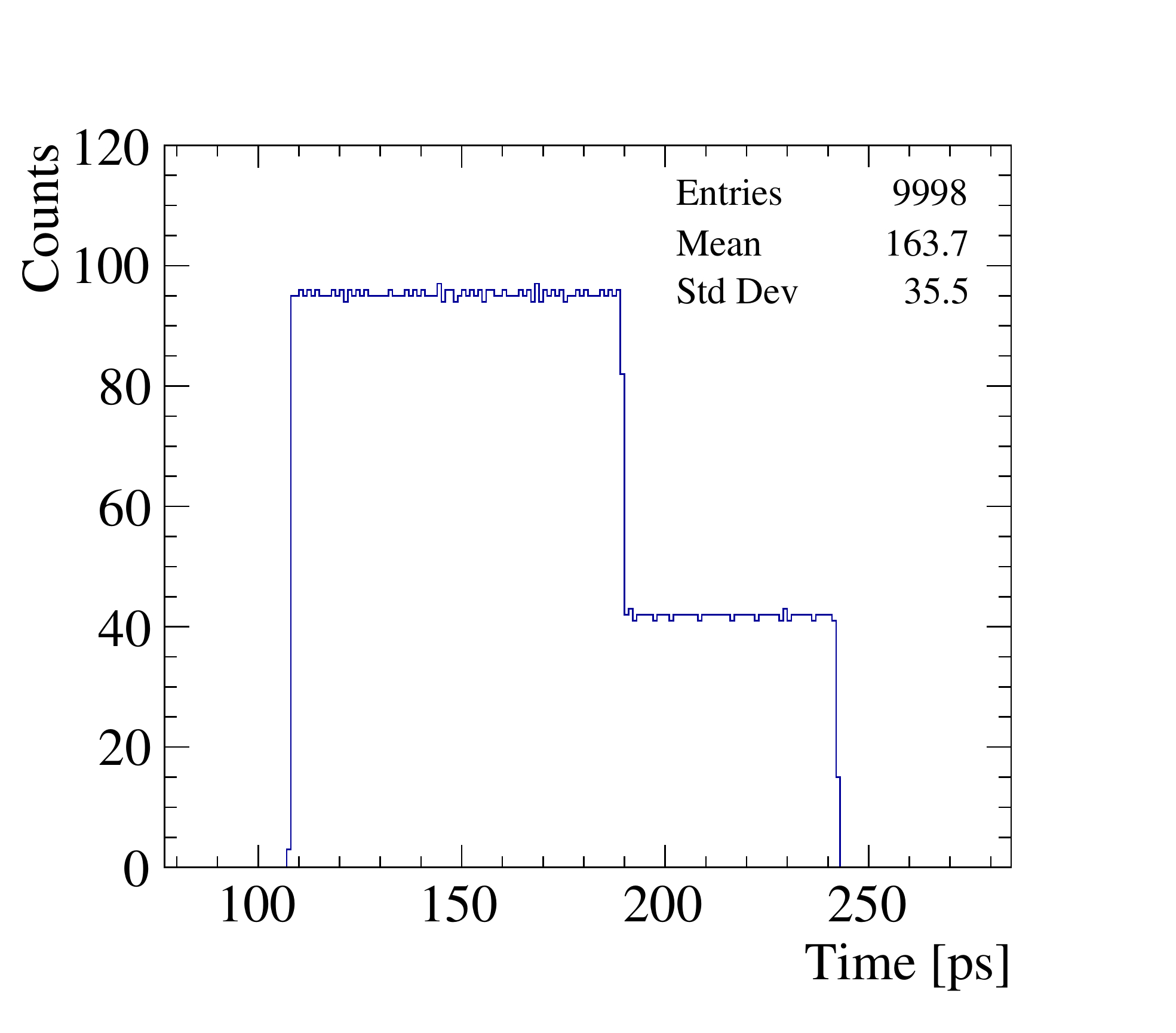}
	\includegraphics[scale=0.274]{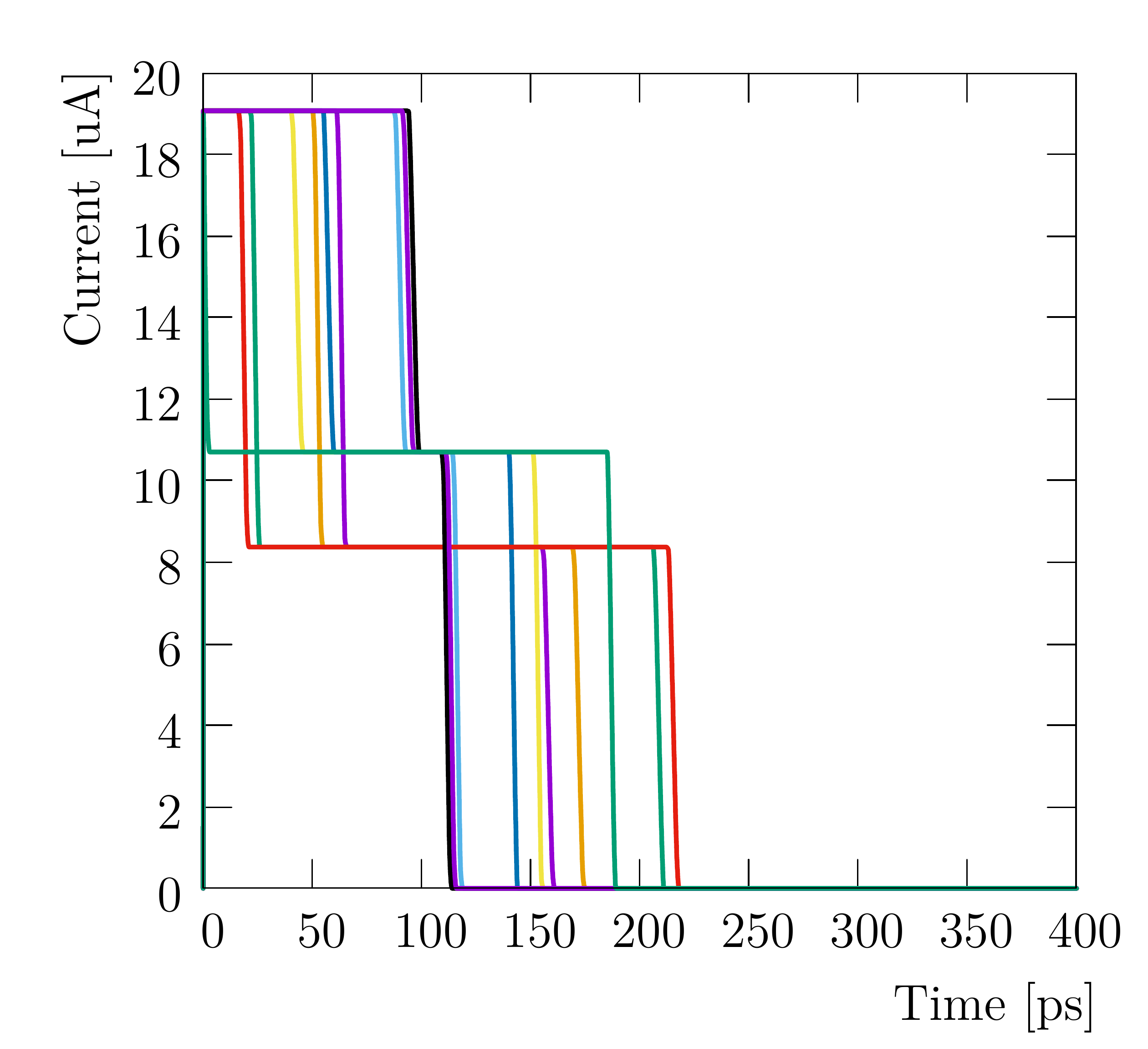}\vspace{-0.5cm}
	\caption{\footnotesize{  CCT distribution of the current used for the simulations with TFBoost (left). Boxed current of the ideal 3D detector (right).} }
	\label{fig:cct_curr}
\end{figure}

\begin{figure}[h]
	\centering
	\includegraphics[width=\textwidth]{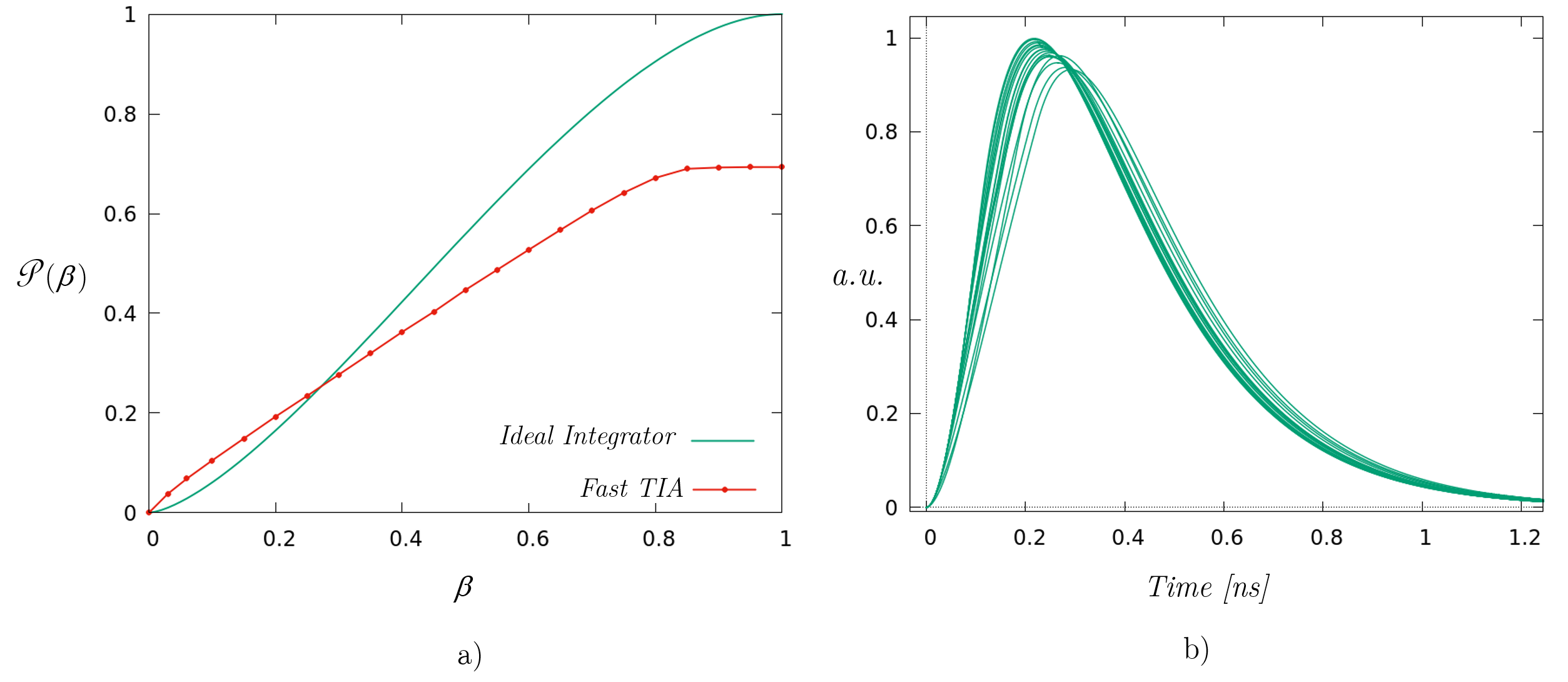}\vspace{-0.5cm}
	\caption{\footnotesize{ a) Comparison of the propagation coefficient relative to the ideal integrator with that obtained with fast TIA electronics. b) Ballistic deficit on the output signals with the fast TIA when $\tau = \overline{t_c} $.} }
	\label{fig:prop_real}
\end{figure}

The parameters to obtain the transient currents (Fig.~\ref{fig:cct_curr} (b)) used for the simulations are chosen in order to represent a realistic 3D-trench silicon sensor:  electrode distance $d = 20~\mu\text{m}$, uniform electric field, both carriers in the condition of saturated drift velocity with  $v_e = 0.107~\mu\text{m/ps}$ and $v_h = 0.0837~\mu\text{m/ps}$. The corresponding CCT distribution is shown in Fig.~\ref{fig:cct_curr} (a). Using Eq.~\ref{sigmatc_vevh} we find the standard deviation of the CCT distribution equal to $\sigma_{t_c}=35.4~\text{ps}$, in agreement with the value found in the simulation, and an average current duration of $~\overline{t_c}\sim 160~\text{ps}$. Since the ratio between the saturation velocities is equal to $k = 0.78$, looking at Fig.~\ref{fig:pdik} we can estimate the time resolution obtainable with slow electronics simply by considering half of $\sigma_{t_c}$, i.e. $\sim17.7~\text{ps}$. With fast electronics the time resolution will depend on the chosen discrimination method and threshold used, thus a scan in the $\beta$ value can be studied by applying a constant fraction discrimination to the simulated output signals. The corresponding propagation coefficient is shown in Fig.~\ref{fig:prop_real} (a).

As can be seen from Fig.~\ref{fig:prop_real} (b), the output signals with fast electronics are characterized by different shapes, since they reflect the different transient current shapes at electronics input. In contrast, in the case of slow electronics the output signal shapes are all equal, being the impulsive response of the electronics.  Fig.~\ref{fig:prop_real} (b) shows the output signals obtained with the convolution of the transfer function in Eq. \ref{eq:RmFast} with the input currents. Having chosen the time constant $\tau$ equal to the average current duration $\overline{t_c}$, for some signals the input charge is not completely integrated, which leads to the well-known phenomenon of ballistic deficit. From the different rise-time values of the signals in Fig.~\ref{fig:prop_real} (b) it is clear that the dispersion of the TOA is maximum when using higher thresholds while it tends to zero as the threshold approaches the signal baseline. Considering a threshold $\beta=0.2$ we obtain a propagation coefficient $\mathscr{P} \sim 0.2$ and consequently an intrinsic temporal resolution of $\sigma_{t_s}=\mathscr{P}(\beta)\sigma_{t_c} \sim 7 ps.$

The real threshold that can be used will obviously depend on the amount of noise present in the system. This will affect both the intrinsic resolution obtainable and another fundamental contribution to the temporal resolution of the system which is given by the electronic jitter $\sigma_{ej}$, which will be discussed in Sec~\ref{subsec:noise}.

\subsection{Intrinsic asymmetry and synchronous region}
\label{subsec:asymTIA}
\begin{figure}[h]
	\centering
	\includegraphics[scale=0.30]{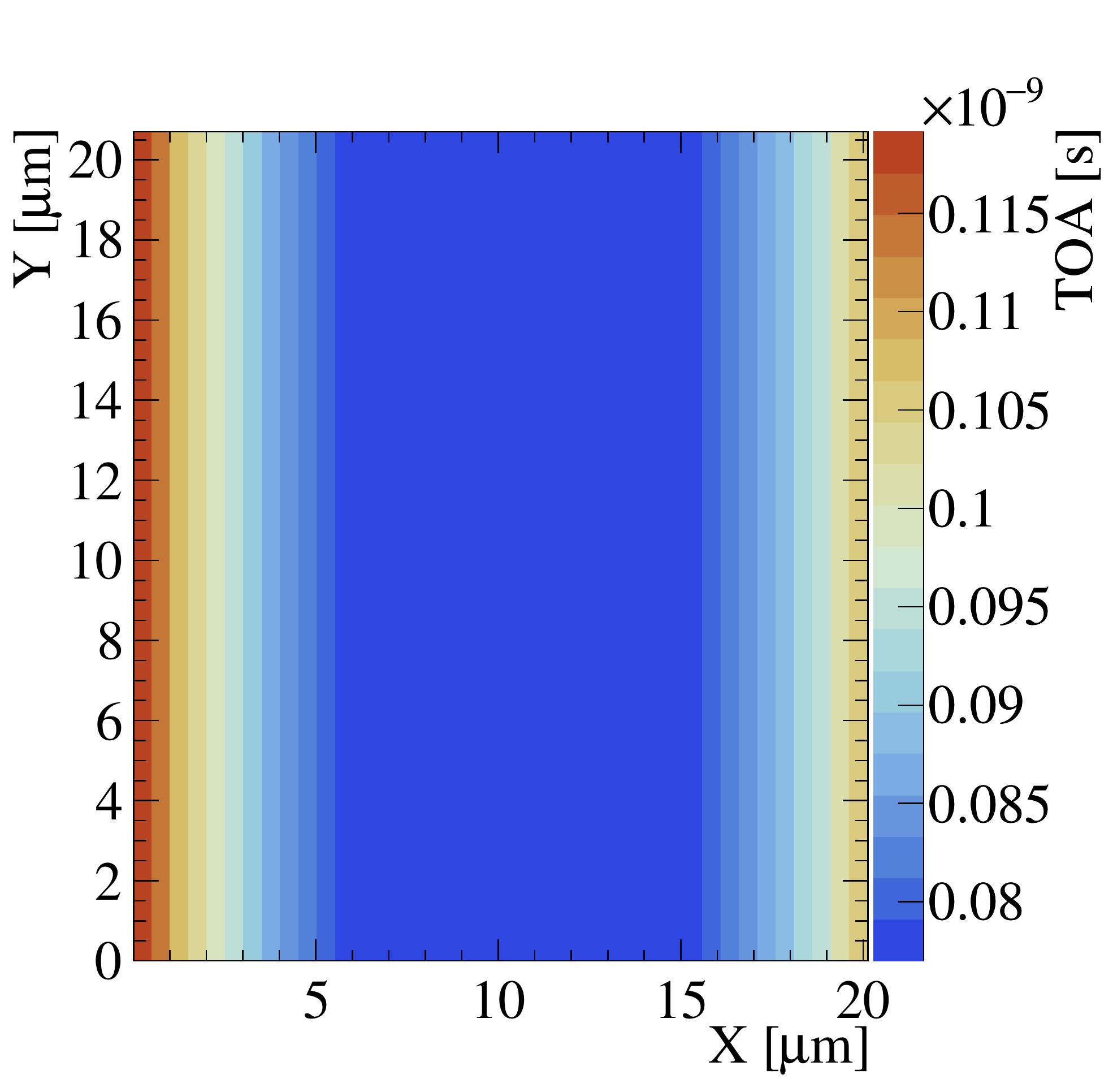}~~~~~
	\includegraphics[scale=0.30]{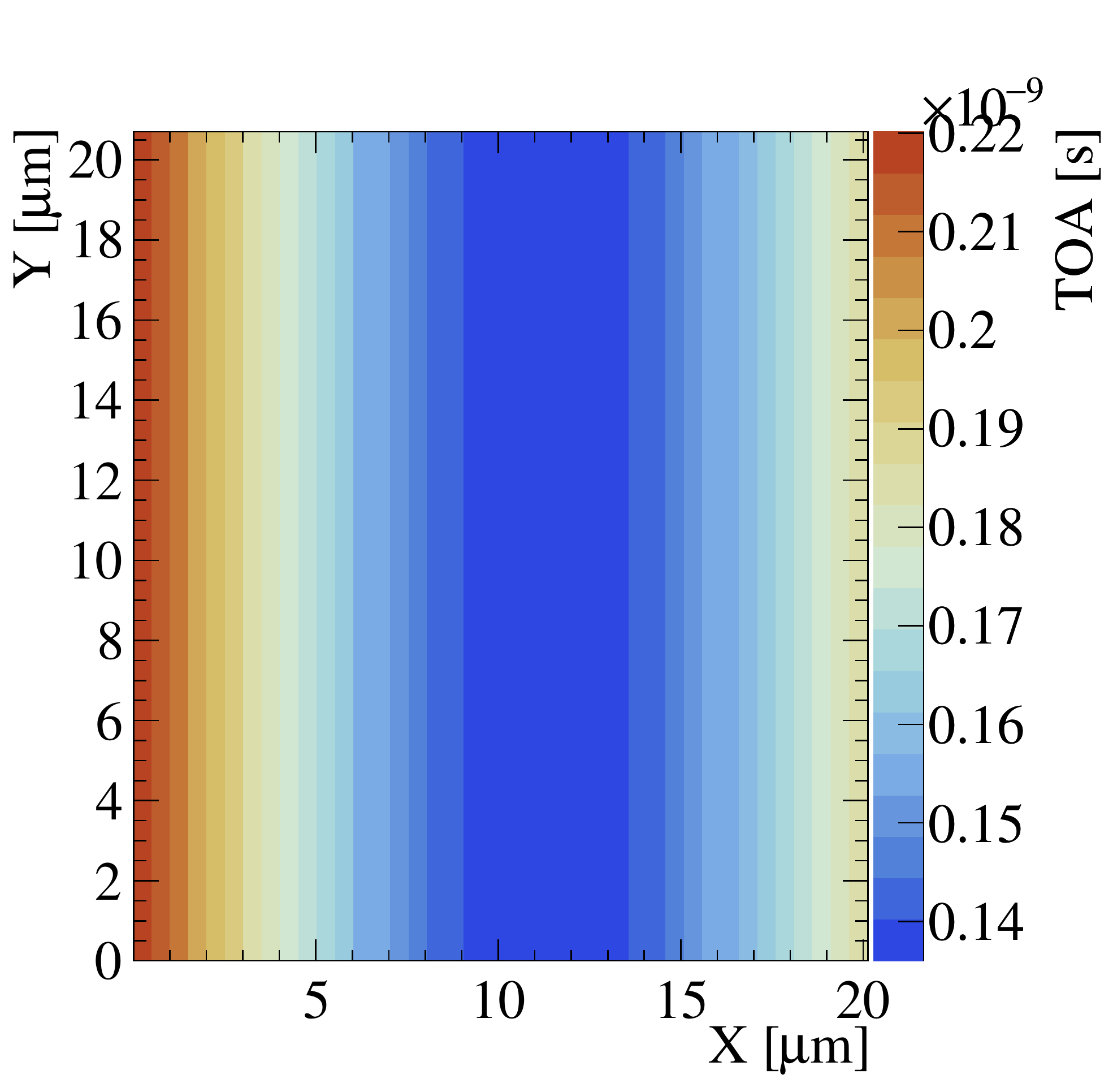}\vspace{-0.4cm}
	\caption{TOA as a function of the impact point in the sensor, obtained by using CFD with $\beta=0.35$ (left) and $\beta=0.8$ (right).}
	\label{fig:tia_synchregion}
\end{figure}

\begin{figure}
    \centering
    \includegraphics[scale=0.66]{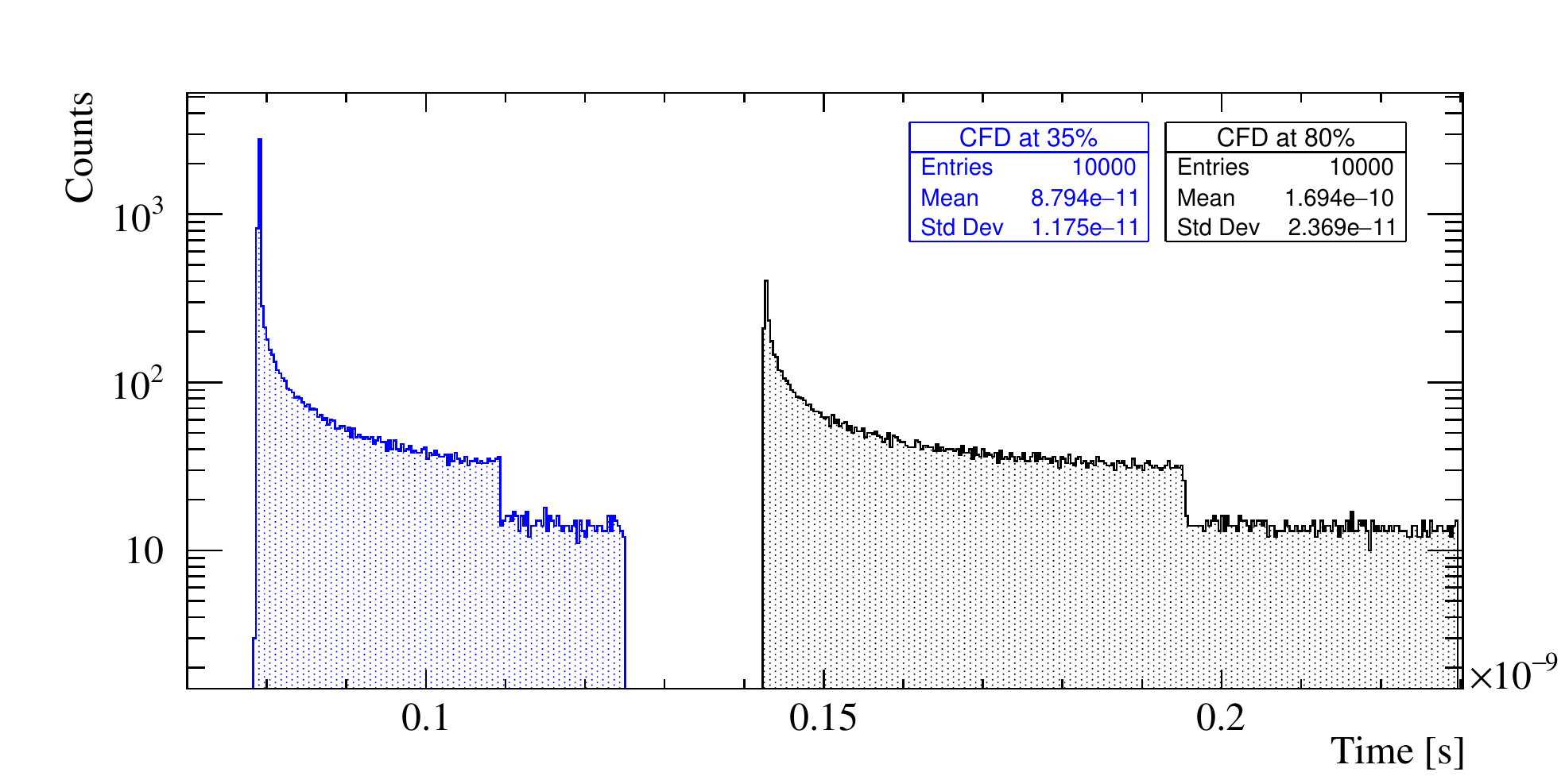}\vspace{-0.2cm}
    \caption{Intrinsic TOA distribution obtained by using CFD with $\beta=0.35$ (blue) and $\beta=0.8$ (black).}
    \label{fig:tia_synchregion2}
\end{figure}

The output signals obtained with the fast TIA have been discriminated with a CFD at different thresholds by using the TFBoost package. The TOA distribution map in the case of two examples, with $\beta=0.35$ and $0.8$, are shown in Fig~\ref{fig:tia_synchregion}, where it is possible to see how the synchronous region is wider at a lower threshold, covering almost half of the whole area of the sensor, while it is tighter when the higher threshold is used. Moreover, in the case of $\beta=0.8$, the TOA distribution is shifted towards higher time values, as shown clearly in Fig~\ref{fig:tia_synchregion2}. 

\begin{figure}[h]
	\centering
	\includegraphics[scale=0.34]{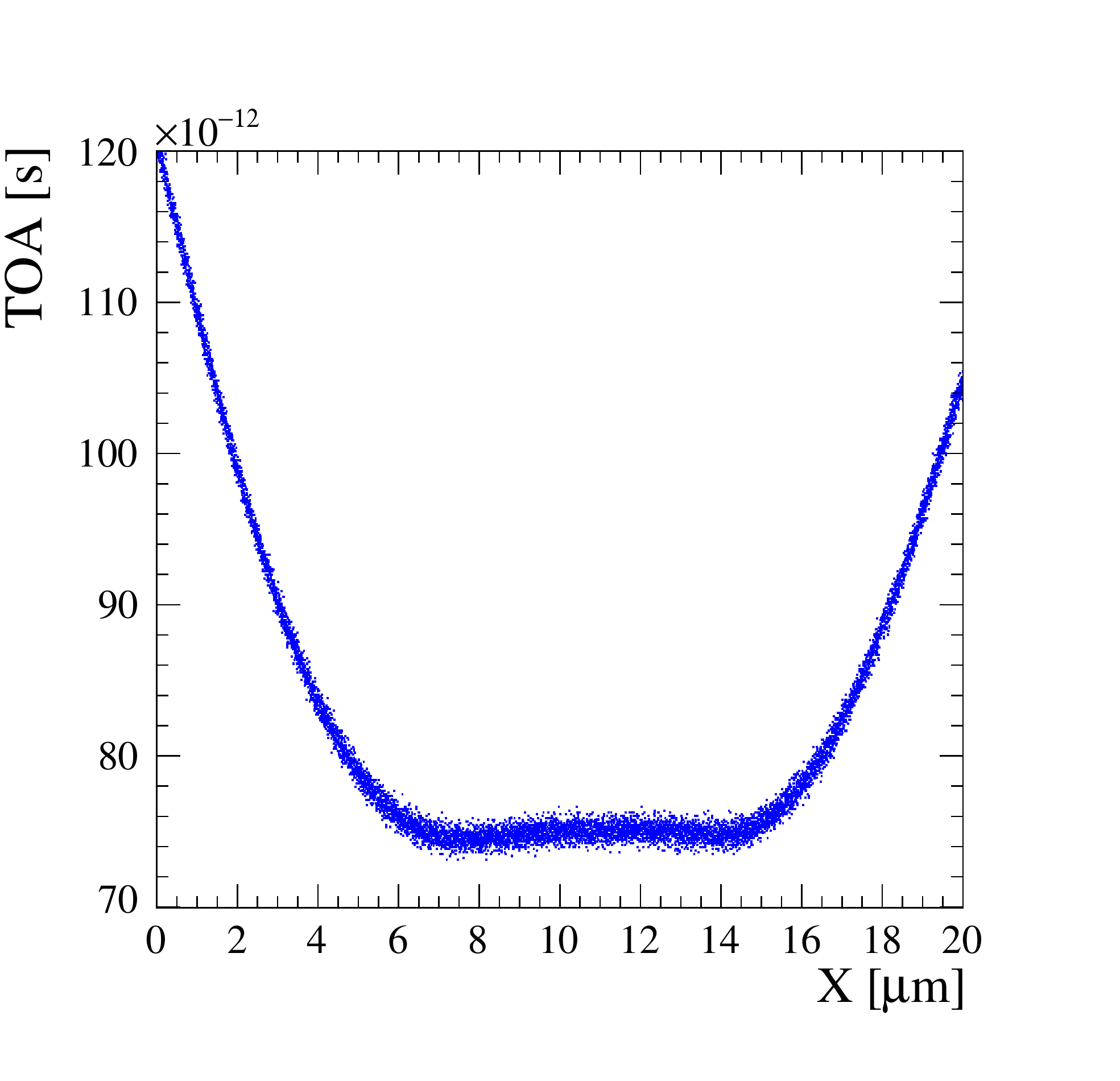}
	\includegraphics[scale=0.34]{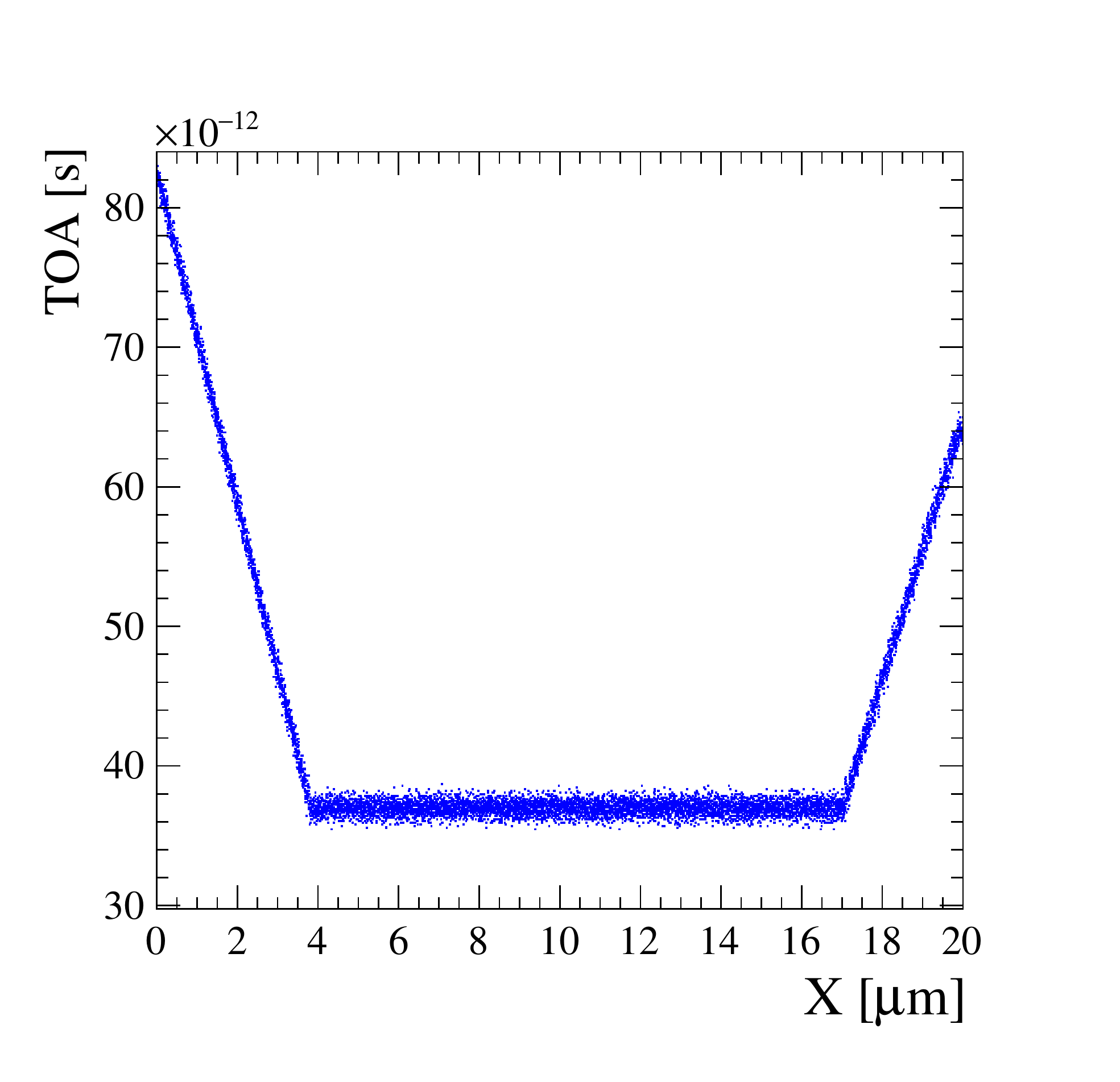}
	\caption{TOA as a function of the impact point in the sensor along the X axis, obtained by using CFD with $\beta=0.35$ with the Ideal Integrator (right) and with the Fast TIA (left).}
	\label{fig:tia_proj}
\end{figure}

This result is compatible with the analytical results obtained in the case of the ideal integrator model, shown in Fig.~\ref{fig:regsinc}. However, an interesting difference to note concerns the shape of the synchronous peak, which is not perfectly sharp as in the case of the ideal integrator, but has a smoother behaviour for increasing time values, as shown in Fig.~\ref{fig:tia_synchregion2} where a log scale for the vertical axis is used to emphasize this effect. Being the transition between the two components smoother, also the edge of the synchronous region is no longer abrupt, as can be seen also in Fig.~\ref{fig:tia_proj}, where the TOA as a function of the particle impact point along the X axis is shown. 

It is important to point out that also with realistic electronics, such as the fast TIA, we find the asymmetry introduced in Sec. \ref{asymmetry} (see Fig.~\ref{fig:tia_synchregion2}). Since we are using transient currents from a simple 3D sensor model, where only the 3D geometry is simulated, the asymmetry of the TOA distribution is due to the intrinsic 3D geometrical properties only, and not due to higher-order effects.

\subsection{Mixture of comprehensive TOA distributions}
\label{subsec:mixturenoise}
\begin{figure}[h]
    \centering
    \includegraphics[scale=0.45]{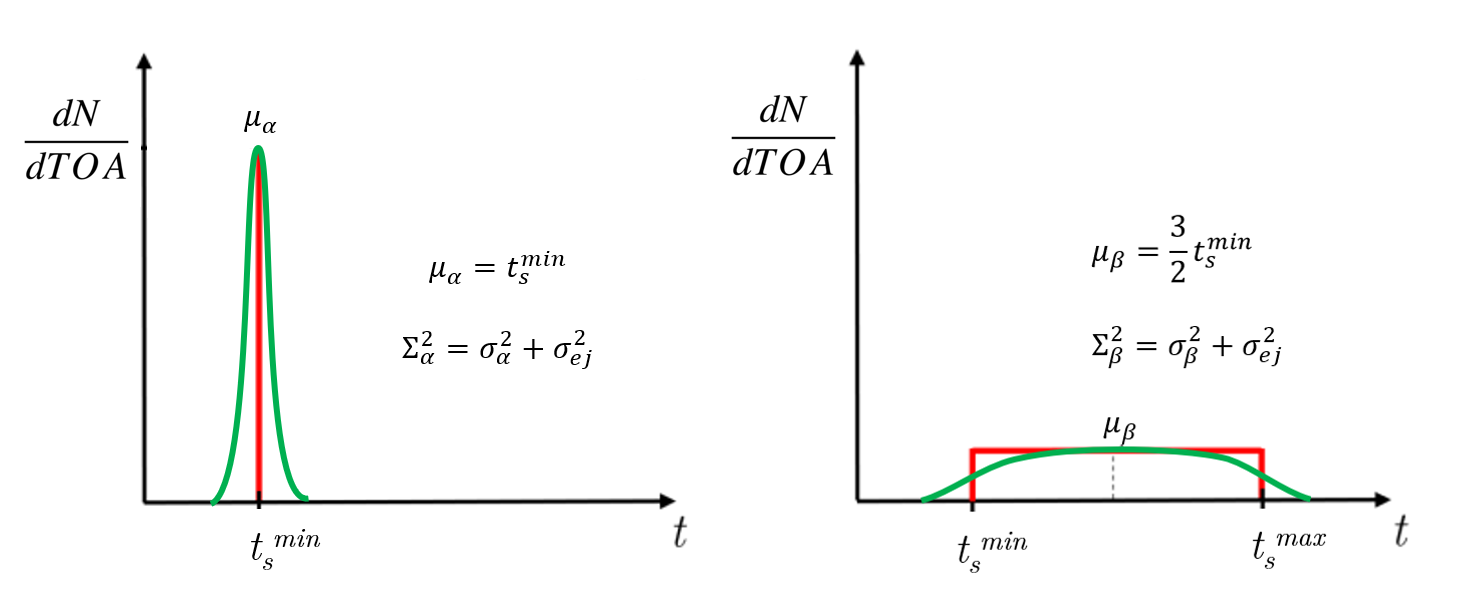}
    \caption{Effect of the electronic jitter on the TOA distribution of the ideal integrator.}
    \label{fig:distr_jitter1}
\end{figure}

\begin{figure}[h]
    \centering
    \includegraphics[scale=0.40]{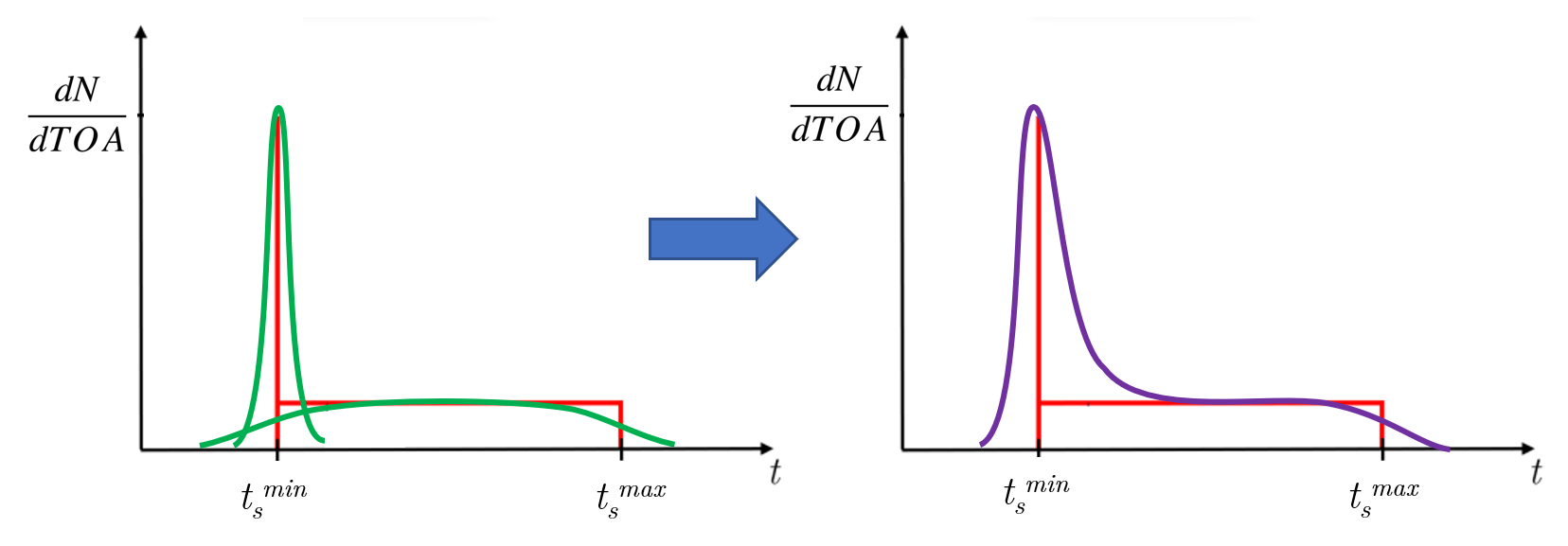}
    \caption{PDF P(t) as a mixture of the distributions $P_\alpha (t)$ and $P_\beta (t)$.}
    \label{fig:distr_jitter2}
\end{figure}

In the presence of electronics noise the intrinsic TOA distribution of the ideal integrator will be distorted and enlarged (Fig.~\ref{fig:distr_jitter1}).
From a mathematical point of view the intrinsic TOA PDF, $p(t)$, is convoluted with the jitter PDF, $J(t)$ with standard deviation $\sigma_{ej}$, leading to the final PDF,
\begin{equation}\label{mixturejitter}
P(t) = p(t) \otimes  J(t) .
\end{equation}
Recalling the mixture distribution in Eq.~\ref{mixturegeneral} we get
\begin{equation}\label{P(t)}
P(t) = \alpha \Big[ p_\alpha(t) \otimes  J(t) \Big ] + \beta \Big[ p_\beta(t) \otimes  J(t) \Big ].
\end{equation}
In order to simplify the notation we indicate the two convoluted PDFs for the synchronous and not synchronous regions as $P_\alpha (t)$ and $P_\beta (t)$ respectively, and their standard deviation as $\Sigma_{\alpha}$ and $\Sigma_{\beta}$ respectively. Being the contributions from the two PDFs uncorrelated, the standard deviation can be summed up in quadrature so that $\Sigma_{\alpha}^2= \sigma_\alpha^2 + \sigma_{ej}^2 $ and $\Sigma_{\beta}^2= \sigma_\beta^2 + \sigma_{ej}^2 $. The total time resolution of the mixture will be then given by
\begin{equation}
\sigma_{t}^2=  \alpha(\Sigma_{\alpha}^2+\mu_{\alpha}^2) + \beta(\Sigma_{\beta}^2+\mu_{\beta}^2) - \mu^2,
\end{equation}
where we made the assumption that the $J(t)$ has its mean $\mu_{J}=0$ so that $\mu_{\alpha}$, $\mu_{\beta}$ and $\mu$ are the same mean values in Eq. \ref{mixture} . An example of the final PDF $P(t)$ is shown in Fig. \ref{fig:distr_jitter2}.

Since the PDF $p_{\alpha}(t)$ has $\sigma_\alpha=0$, we find that the TOA distribution of the ideal integrator has the important characteristic that the synchronous peak relative to the fraction of events $\alpha$, in the presence of electronic jitter, represents the distribution of the jitter itself, being the convolution of $J(t)$ PDF and the synchronous delta-function. However, in the case of fast TIA we found that the synchronous peak, indicated here by $p_\alpha(t)$, is no more a delta-function, as shown in  Fig.~\ref{fig:tia_synchregion2}. Thus, for a real fast TIA the peak function $P_\alpha$ receives contributions from both the jitter and the intrinsic sensor distribution.

\begin{figure}[h] 
	\centering
    \begin{minipage}{.5\textwidth}
        \centering
        \vspace{-6mm}
        \includegraphics[scale=0.39]{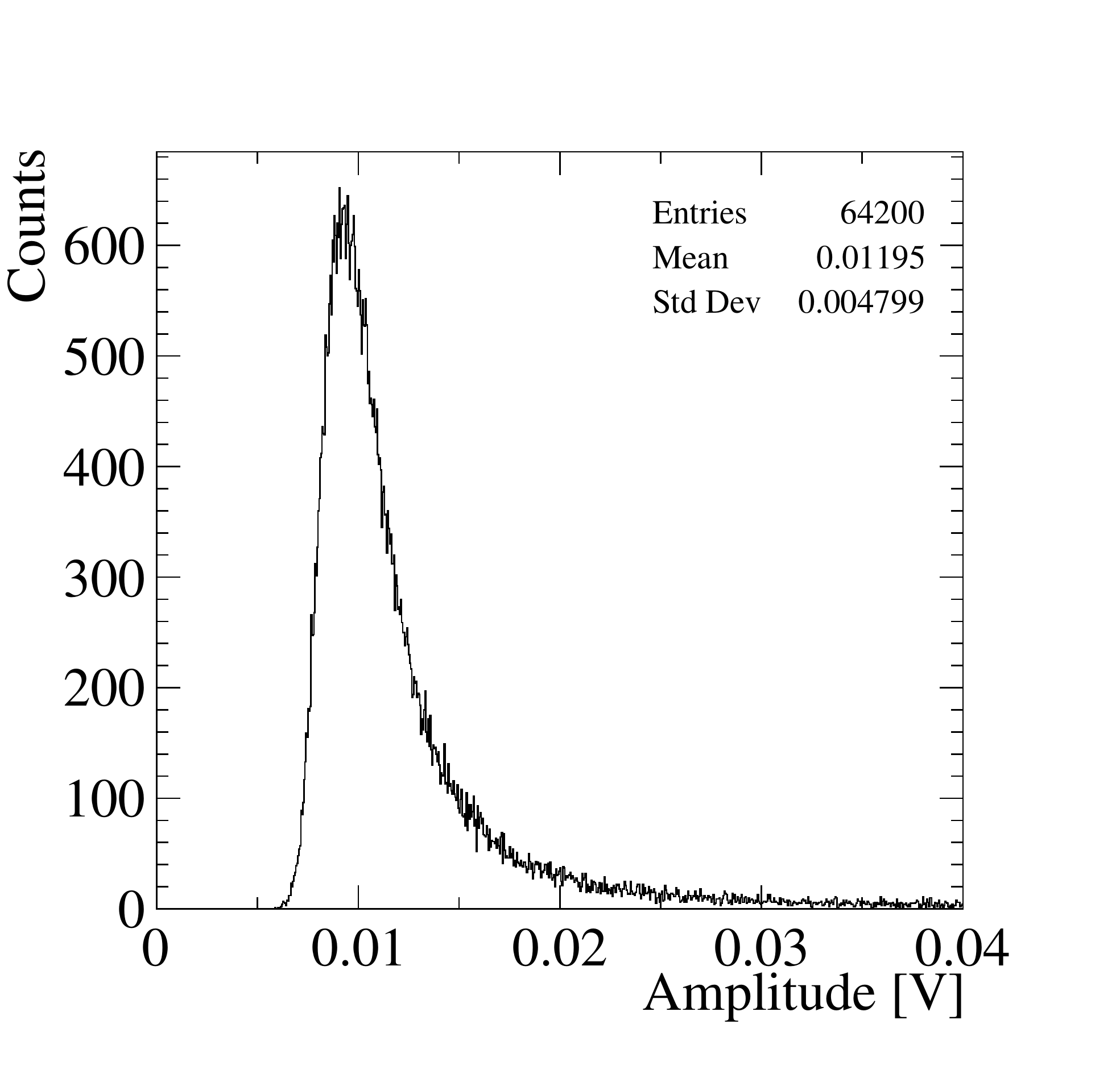}
        \label{fig:prob1_6_2}
    \end{minipage}%
    \begin{minipage}{0.5\textwidth}
        \centering
        \includegraphics[scale=0.38]{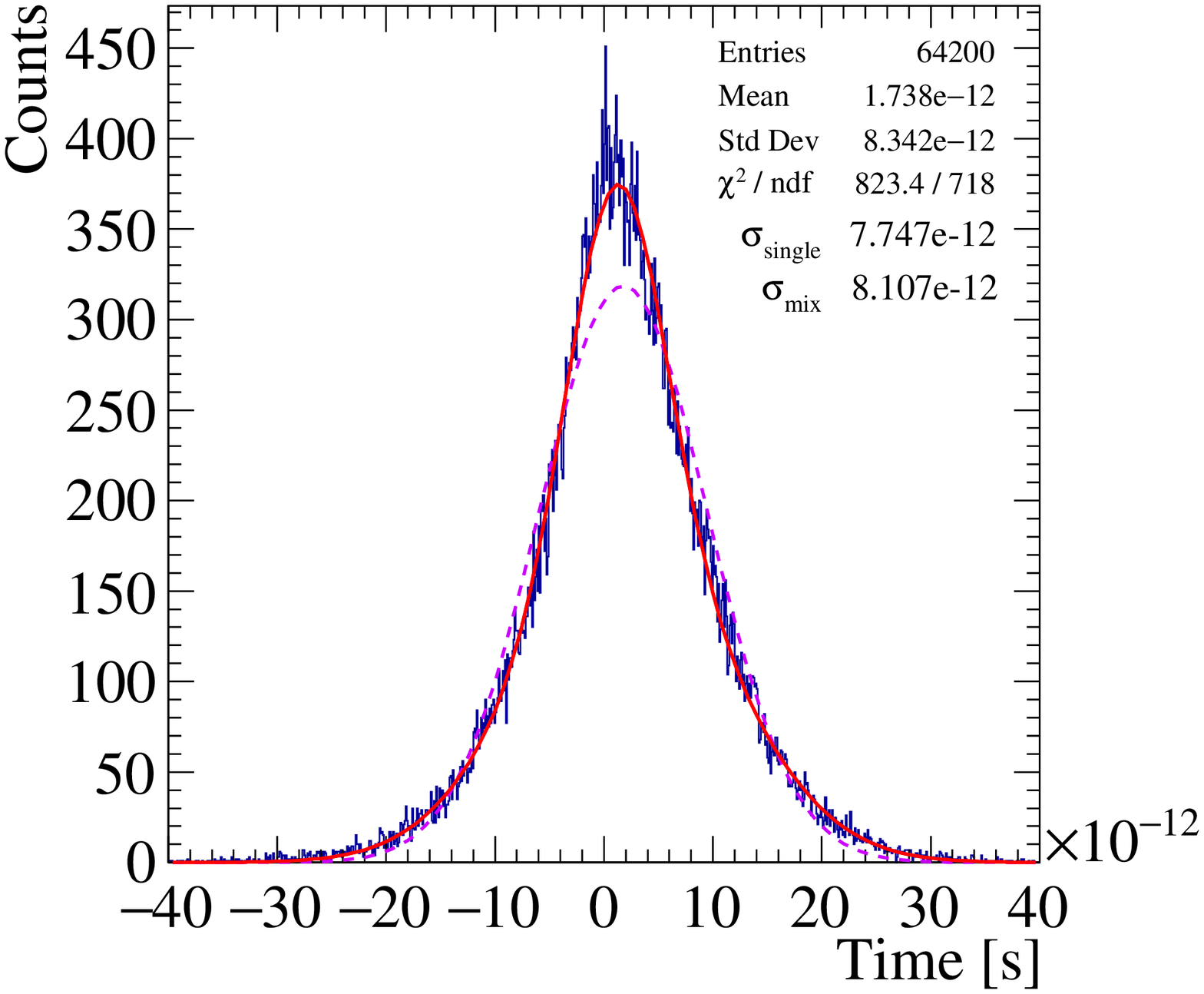}
        \label{fig:prob1_6_1}
    \end{minipage}
    \vspace{-13mm}
	\caption{Left: signal amplitude distribution considering a sensor thickness of 150 $\mu$m obtained with the fast electronics. Right: electronics jitter distribution, obtained as the difference between the TOA measured with and without noise (event-by-event) in the simulation. The solid line represent a fit using the mixture of two Gaussians while the dashed line is the result of a fit with a single Gaussian function.}
	\label{fig:jitter}
\end{figure}

The $P(t)$ function is able to describe experimental data in the case of TOA measurements with 3D-trench sensors and realistic electronics, so it is important to study its analytical properties in order to fit the data accurately. For this purpose, the shape of $J(t)$ plays a crucial role. If the voltage signal shapes and amplitude do not vary significantly event-by-event and if the voltage noise is Gaussian, the electronics jitter $J(t)$ is itself Gaussian, with a standard deviation $\sigma_{ej}$ that depends on the output noise RMS, $\sigma_{V}$, of the amplification stage and on the slope $V'_{\text{thr}}$ of the signal at the chosen discrimination threshold:  $\sigma_{ej} = \sigma_{V} / V'_{\text{thr}}$. The function $J(t)$ starts to deviate from a Gaussian function if one of the mentioned hypotheses is not true. In particular, as we observed in Fig.~\ref{fig:prop_real} (right), fast front-end electronics can be sensitive to the different transient current shapes and be affected by the ballistic deficit. Each different shapes contribute to the jitter function with a different $V'_{\text{thr}}$, thus the final $J(t)$ is no more exactly Gaussian. Another example is when we are in the presence of Landau fluctuations. The signals are characterized by a wide range of amplitudes and, consequently, by a wide range of different slopes $V'_{\text{thr}}$. In order to study these effects we performed simulations with the TFBoost package where we emulate realistic noise samples starting from experimental measurements \cite{Brundu_2021} (in particular we used a Gaussian red noise with 0.985 correlation), and we introduced Landau fluctuations considering the toy model 150 $\mu$m thick 3D sensor Fig.~\ref{fig:jitter} (left). 
\begin{figure}[h]
	\centering
	\includegraphics[scale=0.40]{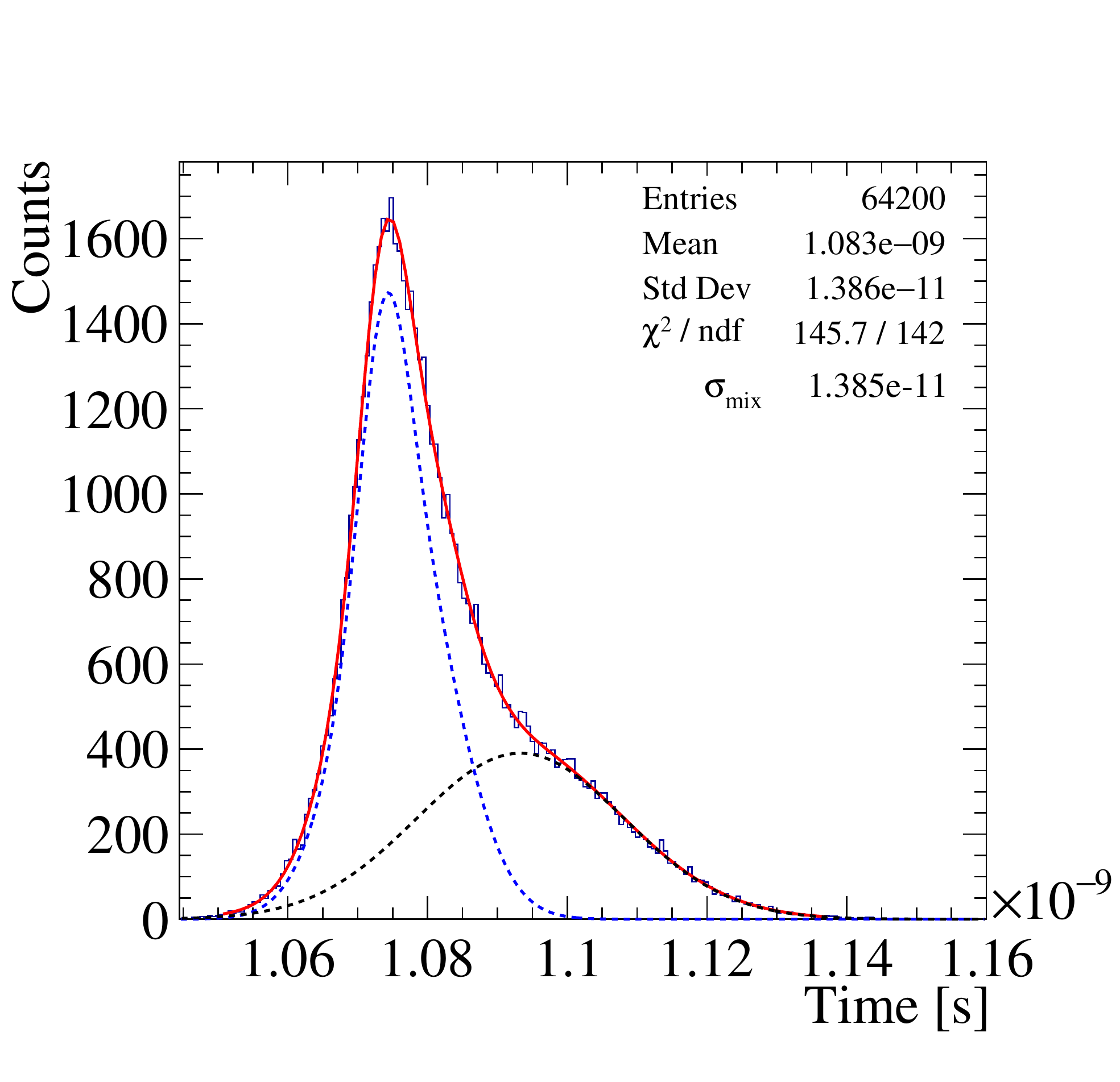}\vspace{-0.4cm}
	\caption{TOA distributions of voltage signals using the fast electronics, $\beta=0.35$ and SNR$\sim 20$. The blue PDF is the mixture of the two Gaussians used to model the synchronous peak, while the black PDF is a single Gaussian to model the not synchronous contribution.}
	\label{fig:toa_final}
\end{figure}

The corresponding $J(t)$ distribution is obtained by calculating the event-by-event difference of TOA values, measured with and without noise, and it is shown in Fig.~\ref{fig:jitter} (right), where the deviation from Gaussianity is evident. In this example the TOA are evaluated with a CFD with $\beta=0.35$, and the threshold is kept fixed between the measurements with and without noise.  
The $J(t)$ can be modelled as a mixture of Gaussian PDFs with different standard deviation $\sigma_{ej}$, corresponding to the different slopes $V'_{\text{thr}}$.  From a pure experimental point of view, it is interesting to investigate how many Gaussian functions are needed within the mixture model to fit realistic data distributions with acceptable accuracy. By using one single Gaussian the systematic uncertainty on the $\sigma_{ej}$ evaluation is up to $\sim 10\%$, while by using two Gaussians it becomes $\sim 2\%$ and it is sufficient to properly model the jitter behaviour, as shown in Fig.~\ref{fig:jitter} (right).
The comprehensive distribution, as described in Eq. \ref{P(t)}, is shown in Fig.~\ref{fig:toa_final}, where the intrinsic asymmetry and effect of the electronic jitter $J(t)$ on the shape of the final PDF can be seen. Given the nature of the two original distributions (synchronous and not synchronous) the proper model to fit the $P(t)$ has to be a mixture of PDFs. 

From a pure experimental point of view it is essential to find a simple mixture model. The one shown in Fig.~\ref{fig:toa_final} is composed of three Gaussians, two of them with the same mean to describe the synchronous peak (which we expect is mainly given by the electronic jitter of the system $J(t)$, as discussed previously), while the third is used to model the not synchronous contribution. The standard deviation of the overall mixture model is consistent with the true standard deviation of the distribution. 

\subsection{Noise contribution to the symmetry of TOA distribution}
\label{subsec:noise}

In the previous Section \ref{subsec:mixturenoise}, we have discussed how the jitter distribution $J(t)$ can modify the shape of the final TOA distribution. This effect can be more or less significant depending on the SNR or, equivalently, on the relative value of the standard deviations $\sigma_{t_s}$ and $\sigma_{ej} $ of the two distributions, $p(t)$ and $J(t)$, respectively.
\begin{figure}[h]
	\centering
	\includegraphics[scale=0.30]{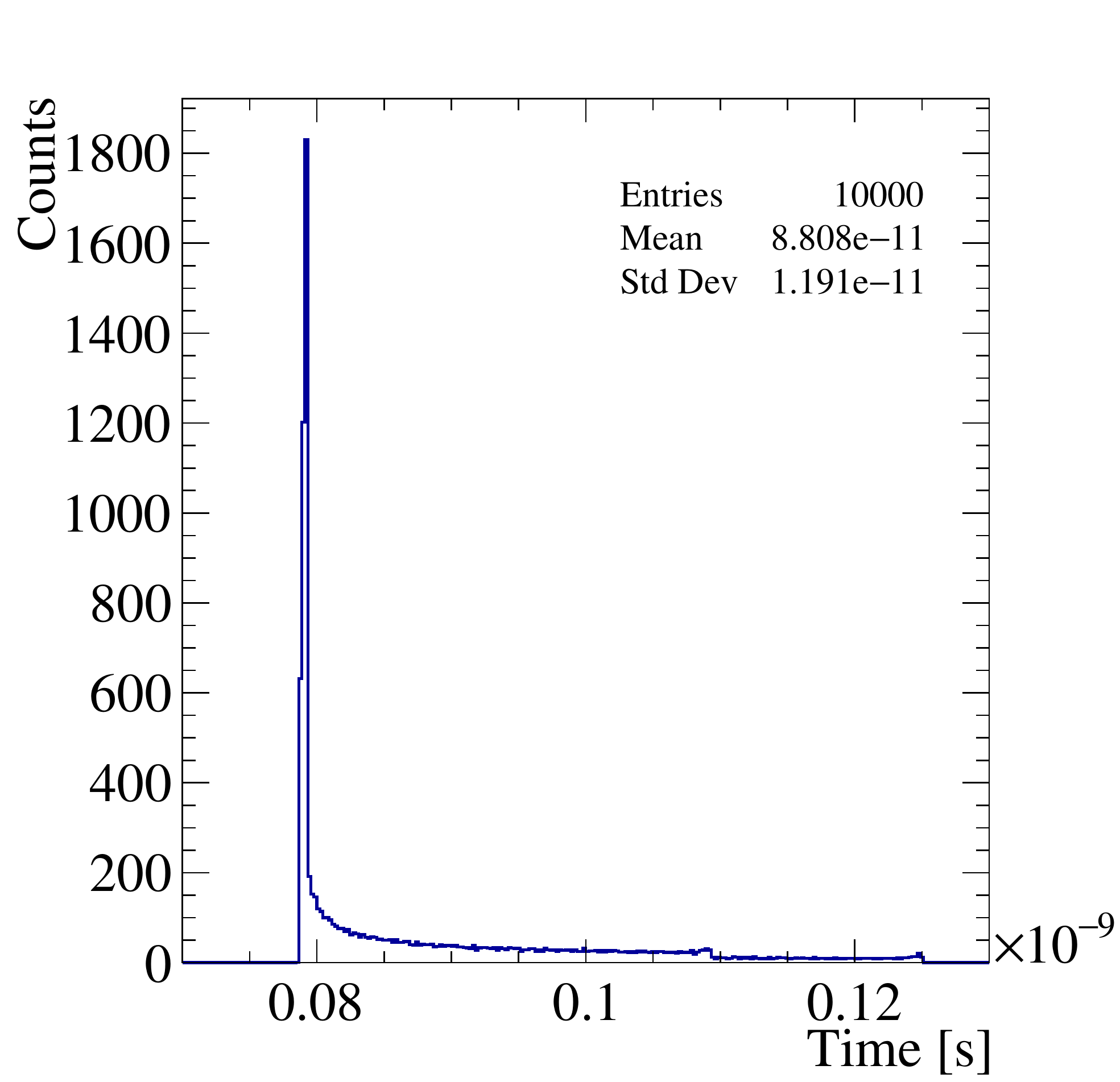}
	\includegraphics[scale=0.30]{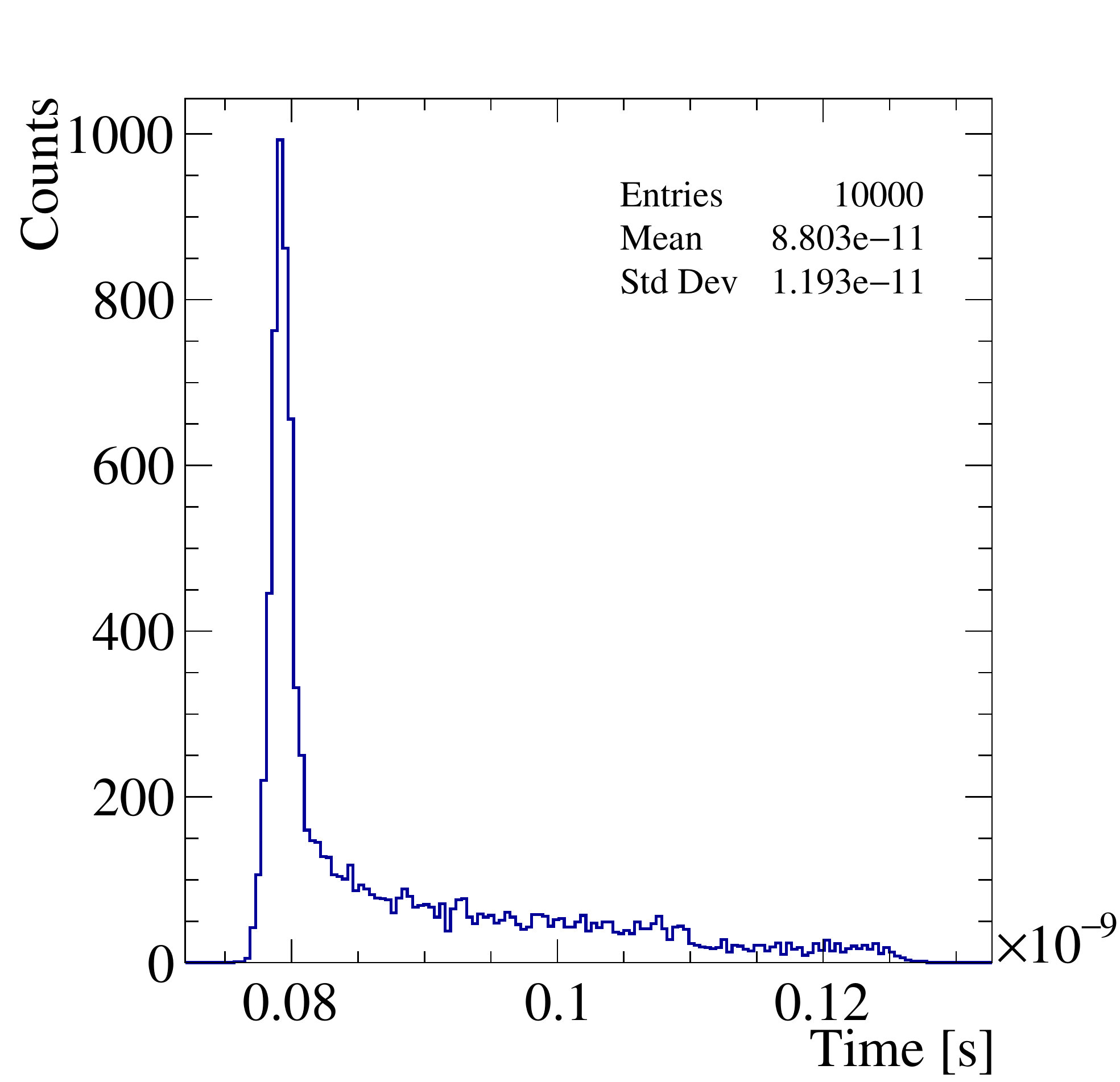}
	\includegraphics[scale=0.30]{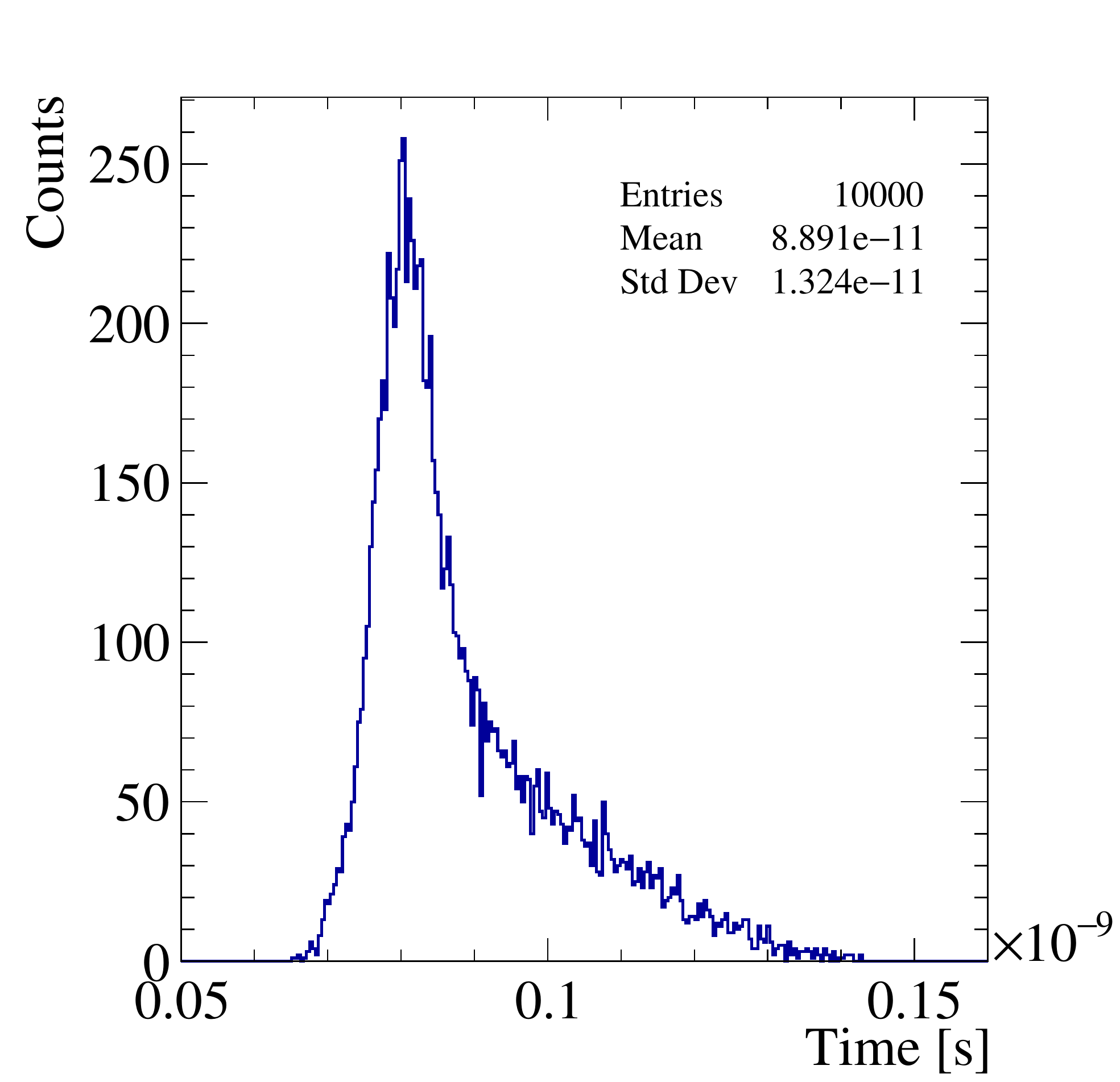}
	\includegraphics[scale=0.30]{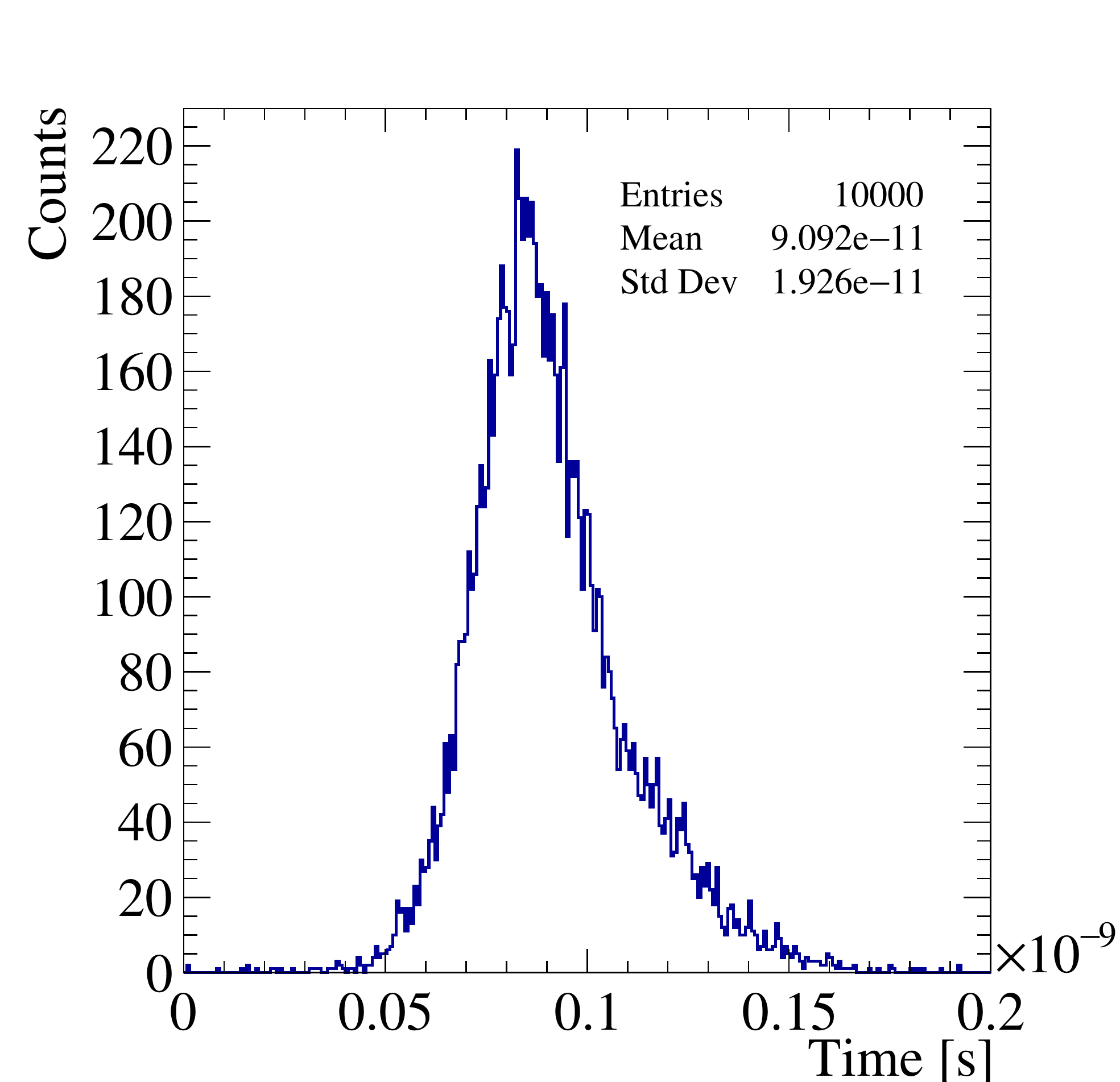}
 \vspace{-3mm}
	\caption{Intrinsic TOA distributions for different SNR: $ \text{SNR} = \infty$ (top left), $ \text{SNR} = 40$ (top right), $ \text{SNR} = 20$ (bottom left), $ \text{SNR} = 10$ (bottom right).}
	\label{fig:tia_noise_asym}
\end{figure}

Fig~\ref{fig:tia_noise_asym} shows the TOA distribution for different SNR values. In the limit of infinite SNR there is no jitter, and we find the intrinsic distribution of the sensor already seen in Fig.~\ref{fig:tia_synchregion2}, while for lower SNR, the TOA distribution becomes more and more symmetrical.
Depending on the amount of jitter present, the mixture model for the final distribution must be adapted to the specific case. For example, looking at Fig~\ref{fig:tia_noise_asym}, for SNR lower than 20, a mixture model with two Gaussian functions is accurate enough to properly describe the final distribution, and allows to estimate the  time resolution with an accuracy of order $\mathcal{O}(1\%)$.

Since the intrinsic asymmetry is a peculiar feature of 3D sensors, it is important to point out that the asymmetry of the comprehensive distribution, which has been brought out with this simple 3D sensor model, experimentally cannot be ignored just by taking into account the resolution of the peaking structure, because this can lead to a severe underestimation of the time resolution $\sigma_t$. Other aspects that affect the duration of induction of the carriers, such as weighting field non-uniformity, will further modify the TOA distribution, whose shape will be however characterised by intrinsic asymmetry.

\section{Conclusions}
In this work the intrinsic timing properties of a single-pixel 3D-trench sensor have been discussed in the case of fast front-end electronics.
In order to perform the study, the transient currents have been obtained analytically by using a simple ideal model to take into account the 3D geometry of the trench structure. The first result is that we can analytically relate the time resolution, i.e. the standard deviation of TOA distribution, $\sigma_{t_s}$, and the CCT standard deviation $\sigma_{t_c}$ through the timing propagation coefficient $\mathscr{P}$, using an ideal integrator as fast electronics. In this case, the propagation coefficient $\mathscr{P}$ depends on the discrimination threshold, in contrast to the slow electronics case. Moreover, we found that the TOA distribution of the single-pixel 3D-trench sensor is composed of two contributions: a synchronous component ideally with $\sigma_{t}=0$, corresponding to all the events given by the particles passing through the \textit{synchronous region} of the pixel active volume, and a non-synchronous component. 
This property leads to a strong asymmetry of the TOA distribution even considering non-ideal front-end electronics such as fast TIA. This \textit{intrinsic asymmetry} is clearly mitigated by the presence of noise, but for SNR $>10$ it remains visible in the final TOA distribution, as already observed in accurate simulations and experimental results \cite{Brundu_2021, Lampis:2022lpj}. We remark that the presence of the synchronous region and the corresponding TOA intrinsic asymmetry arises from the 3D-trench geometry only and it represents a fundamental feature of these sensors. Furthermore we proposed, as an experimental procedure, the usage of a probabilistic mixture model in order to properly describe the TOA contributions and determine correctly the final time resolution. The usage of different models, in particular the ones neglecting the asymmetric right tail of the distribution, could lead to a wrong estimation of the time resolution.

\bibliographystyle{unsrt} 
\bibliography{bibl}

\end{document}